\documentclass[usedcolumn,useAMS,usenatbib]{mn2e}
\usepackage{times,rotating}
\usepackage{graphicx,subfigure,epsf}

\def\xmm{{\it XMM-Newton\/}}
\def\cha{{\it Chandra\/}}
\def\etal{et al.\ }

\def \h50 {$h_{50}$}
\def \h70 {$h_{70}$}

\def \nh {$N_{\rm H}$}

\begin{document}

\title[High redshift FRII radio galaxies: X-ray properties of the
cores]{High redshift FRII radio galaxies: X-ray properties of the cores}
\author[E. Belsole \etal]{E. Belsole$^1$\thanks{E-mail:
e.belsole@bristol.ac.uk},  D. M. Worrall$^1$ and M.J. Hardcastle$^2$\\
$^1$ Department of Physics, University of Bristol, Tyndall Avenue,
Bristol BS8 1TL, UK\\
$^2$  School of Physics, Astronomy and Mathematics, University of Hertfordshire, College Lane, Hatfield, Hertfordshire AL10 9AB}

\date{Accepted . Received ; in original form }
\maketitle

\begin{abstract}
We present an extensive X-ray spectral analysis of the cores  of 19
FRII sources in the redshift range $0.5<z<1.0$ which were selected to be
matched in isotropic radio power. The sample consists
of 10 radio galaxies and 9 quasars. We compare our results with the
expectations from a unification model that ascribes the difference
between these two types of sources to the viewing angle to the line of
sight, beaming and
the presence of a dust and gas torus.
We find that the spectrum of all the quasars can be fitted
with a single power law, and that the
spectral index flattens with decreasing angle to the line of sight. We
interpret this as the effect of increasingly dominant inverse
Compton X-ray emission, beamed such that the jet emission outshines
other  core components.
For up to 70 per cent of the radio galaxies we detect intrinsic
absorption; their core spectra are best fitted with an unabsorbed steep 
power law of average spectral index $\Gamma=2.1$ and an absorbed  power law of
spectral index $\Gamma=1.6$, which is flatter than that observed for
radio-quiet quasars.  
We further conclude that the presence of a jet affects the spectral
  properties of absorbed nuclear emission in AGN. In radio galaxies,
  any steep-spectrum component of nuclear X-ray emission, similar to
  that seen in radio-quiet quasars, must be masked by a jet or by 
  jet-related emission.
\end{abstract}


\section{Introduction}\label{intro}

Active Galactic Nuclei (AGN) are observed across the whole electromagnetic
spectrum, but  X-ray observations probe the process of converting
gravitational energy into radiation most directly, as spectral and
variability studies have shown that at least some of the X-ray emission comes from regions very close
to the central engine.
The most luminous AGN are found at high redshift and thus  their
study probes accretion processes and AGN environment early in the Universe.

AGN comprise a large variety of objects, for
which both simple and more complicated classification  schemes have been
proposed. Unification Models explain the observed
differences among AGN as the result of orientation with respect to
the line of sight (e.g. \citealt{barthel89, UrryPadovani95}). In
particular, it is proposed that  powerful radio galaxies and
radio-loud quasars
are the same objects, and an obscuring torus is invoked to hide the
nucleus of objects (namely radio galaxies) viewed at large
angles to the jet axis (e.g. \citealt{barthel89}).

Large orientation-unbiased samples of AGN are needed to test
unification. They must be observed at many wavelengths, in order to
detect both isotropic and 
anisotropic emission. At low radio frequencies optically-thin synchrotron radiation from the radio
lobes of radio-loud AGN
dominates the emission, and this lobe emission should be isotropic.
Thus, selection of objects via their 
low-frequency radio emission represents the most reliable  method for
selecting an orientation-unbiased sample. The low-frequency radio
emission is also an indirect measure of the power 
supplied by the jet over the lifetime  of the source (some 10$^8$ yr). 

X-ray emission on small spatial scales, particularly at soft energies,
is expected to be anisotropic because of the obscuring torus
required in unified models to obscure the broad-line region and
  optical continuum emission. For those sources which are radio-loud,
 high-frequency nuclear radio emission, probing sub-arcsecond scales
of the source, is  explained as synchrotron radiation  from the unresolved
bases of  relativistic jets, which
is anisotropic due to relativistic beaming. 
The  correlation found using {\it ROSAT} data between the nuclear, soft X-ray emission and the
core radio emission of powerful radio-loud AGN (e.g. \citealt{mjh99}, \citealt*{brinkmann97}) thus implies that at least part of the 
 soft X-ray emission is also relativistically beamed and originates at
the base of the jet \citep{mjh99}. From {\it ROSAT} data it had
  earlier been
found  that lobe-dominated quasars tend to lie above the flux-flux
correlation valid for core-dominated quasars  \citep{dmw94},
supporting the idea that lobe-dominated quasars are viewed at an angle to
the line of sight such that the observer sees in X-ray  both  the
jet-dominated component and a more isotropically emitted, probably
nucleus-related component, with the two components being more
  similar in flux density than for the
core-dominated quasars. This suggestion is further supported by the
rather flat X-ray spectra of radio-loud (core-dominated or
blazar-type) quasars, with slopes
$\Gamma\sim1.5-1.6$ (e.g., \citealt{ww90, brinkmann97,reeves97,rt00,gambill03}), which may be compared to the values of $\Gamma\sim2$ more commonly found
in their radio-quiet counterparts (e.g. ,
\citealt{yuan98,reeves97,rt00}). The flat 
 spectrum has been  interpreted as the result of beamed emission from the jet.

X-ray results supporting unification of quasars with powerful, (narrow-line)
radio galaxies (i.e. that those sources are viewed on the plane of the
sky and are thus strongly obscured by a torus), especially at redshift
above 0.5, are less abundant. Few such radio galaxies were detected in
X-ray with {\it ROSAT}.
However, \citet{dmw94} found that the core soft X-ray emission of the
galaxy 3C\,280 lies on an extrapolation of 
the correlation obtained for core-dominated quasars, and interpreted
the result as due to X-ray emission from a jet-related
component, with the nucleus-related component being obscured by a
putative torus. This result was later extended to a a larger
  sample of radio galaxies by \citet{mjh99}. 

In the simple picture of an obscuring torus, quasar light  heats the gas and dust of the torus and 
thermal radiation is re-emitted isotropically in the mid/far-infrared in
order to maintain energy balance in the inner regions. Thus, in
principle, far-infrared radiation should provide an
orientation-independent measure of the emitted power
from the central engine. Studies based on data from {\it IRAS} (\citealt*{heckman92};
\citealt*{hes95}), and later the Infrared Space Observatory ({\it ISO})  (e.g. \citealt{meis01})
 were controversial because of the limited sensitivity and  spectral
coverage that made  it difficult
to separate star formation heating and AGN heating.  The sensitivity in the mid/far-infrared of the {\em Spitzer} satellite provides for the first time
the possibility to test this hypothesis for a large number of
objects.  \citet{shi05} have found that indeed  AGN heating 
contributes more than 50 per cent of the far-infrared luminosity in
most objects from a sample of 20 radio-loud galaxies and quasars.

A study of a  sample of sources at a range of orientations  for which X-ray and mid/far-infrared
properties derived from high-quality data can be compared, is useful for understanding selection effects in X-ray-selected samples and for comparison with the new understanding of
  these sources coming from {\it Spitzer} data. Here we present high-quality X-ray observations obtained with
\cha\ and \xmm\ of 19 sources
in the redshift range 0.5$<z<1.0$, which are mostly part of a larger sample of
Faranoff-Riley type II (FRII) radio galaxies and quasars currently being  observed with {\em
Spitzer}. In this paper we focus on the X-ray properties of the nuclear
regions; we will address  the environment of
these sources (which is also a potential source of isotropic X-ray
emission) in a forthcoming paper.

Throughout this paper we use the concordance cosmology with $h_0 =
H_0/100\ {\rm km}\ {\rm s}^{-1}\ {\rm Mpc}^{-1}
= 0.7$, $\Omega_{\rm M} = 0.3$, $\Omega_{\Lambda} = 0.7$. If not
otherwise stated, errors are quoted at $1\sigma$ for one interesting
parameter.

\begin{table*}
\begin{center}
\caption{The sample}
\label{tab:sources}
\begin{tabular}{l|cclclcr}
\hline
Source & RA(J2000) & Dec(J2000) & redshift  & scale &type & \nh & Comments\\
        & $^{\rm h~m~s}$ &$^{\circ~\prime~\prime\prime}$ & &
        kpc/arcsec & & 10$^{20}$ cm $^{-2}$\\
\hline
3C\,6.1 &  00 16 30.99 & +79 16 50.88  & 0.840 &7.63&  NLRG & 14.80&\\  
3C\,184 &  07 39 24.31 & +70 23 10.74  & 0.994 &8.00&  NLRG & 3.45& \\
3C\,200 &  08 27 25.44 & +29 18 46.51  & 0.458 &5.82&  NLRG & 3.74& \\
3C\,207 &  08 40 47.58 & +13 12 23.37  & 0.684 &7.08&  QSO& 4.12& \\
3C\,220.1& 09 32 39.65 & +79 06 31.53  & 0.610 &6.73&  NLRG & 1.87& \\
3C\,228 &  09 50 10.70 & +14 20 00.07  & 0.552 &6.42&  NLRG & 3.18& no
{\em Spitzer}\\
3C\,254 &  11 14 38.71 & +40 37 20.29  & 0.734 &7.28&  QSO& 1.90& \\
3C\,263 &  11 39 57.03 & +65 47 49.47  & 0.646 &6.90&  QSO& 1.18& \\
3C\,265 &  11 45 28.99 & +31 33 49.43  & 0.811 &7.54&  NLRG & 1.90& no
{\em Spitzer} \\
3C\,275.1& 12 43 57.67 & +16 22 53.22  & 0.557 &6.40&  QSO& 1.99& \\
3C\,280 &  12 56 57.85 & +47 20 20.30  & 0.996 &8.00&  NLRG & 1.13& \\
3C\,292 &  13 50 41.95 & +64 29 35.40  & 0.713 &6.90&  NLRG & 2.17& \\
3C\,309.1& 14 59 07.60 & +71 40 19.89  & 0.904 &7.80&  GPS-QSO& 2.30&\\
3C\,330 &  16 09 34.71 & +65 56 37.40  & 0.549 &6.41&  NLRG & 2.81 & \\
3C\,334 &  16 20 21.85 & +17 36 23.12  & 0.555 &6.38&  QSO & 4.24& \\
3C\,345 &  16 42 58.80 & +39 48 36.85  & 0.594 &6.66&  core-dom
QSO&1.13 & no {\em Spitzer}\\
3C\,380 &  18 29 31.78 & +48 44 46.45  & 0.691 &7.11&  core-dom QSO& 5.67 &\\
3C\,427.1& 21 04 06.38 & +76 33 11.59  & 0.572 &6.49&  LERG&  10.90&\\
3C\,454.3& 22 53 57.76 & +16 08 53.72  & 0.859 &7.68&  core-dom QSO &
6.50 & no {\em Spitzer}\\
\hline
\end{tabular}
\vskip 10pt
\begin{minipage}{13cm}
Galactic column density is from \citet{dlnh}; NRLG means Narrow Line
Radio Galaxy; LERG means low-excitation radio galaxy. Redshifts and
positions are taken from \cite{3crrcat}.
\end{minipage}
\end{center}
\end{table*}

\section{The sample}\label{sec:sample}
The  AGN for our study were chosen from an 
orientation-unbiased parent sample. As discussed above, a reliable method for such a criterion is
selection on the basis of the low-frequency radio emission. The best known and best 
studied orientation-unbiased sample of radio sources is the 3CRR sample
\citep*{3crrcat}, selected at 178 MHz.  There are 50 3CRR  sources at  redshift $0.5<z<1.0$, of which
$\sim 2/3$ are radio galaxies and $\sim 1/3$ are quasars.  We do not aim to
carry out a statistical test of unification models, for which a random selection from the parent
sample would be necessary. Instead we intend to look for
differences in the X-ray emitting components
between quasars and radio galaxies, as would be
expected in unification schemes. For this reason we base our work on a
sub-sample of 34 high-redshift 3CRR radio galaxies and quasars that are
being observed with {\em Spitzer} as part of a guaranteed-time
programme.

\begin{figure}
\includegraphics[scale=0.55,angle=0,keepaspectratio]{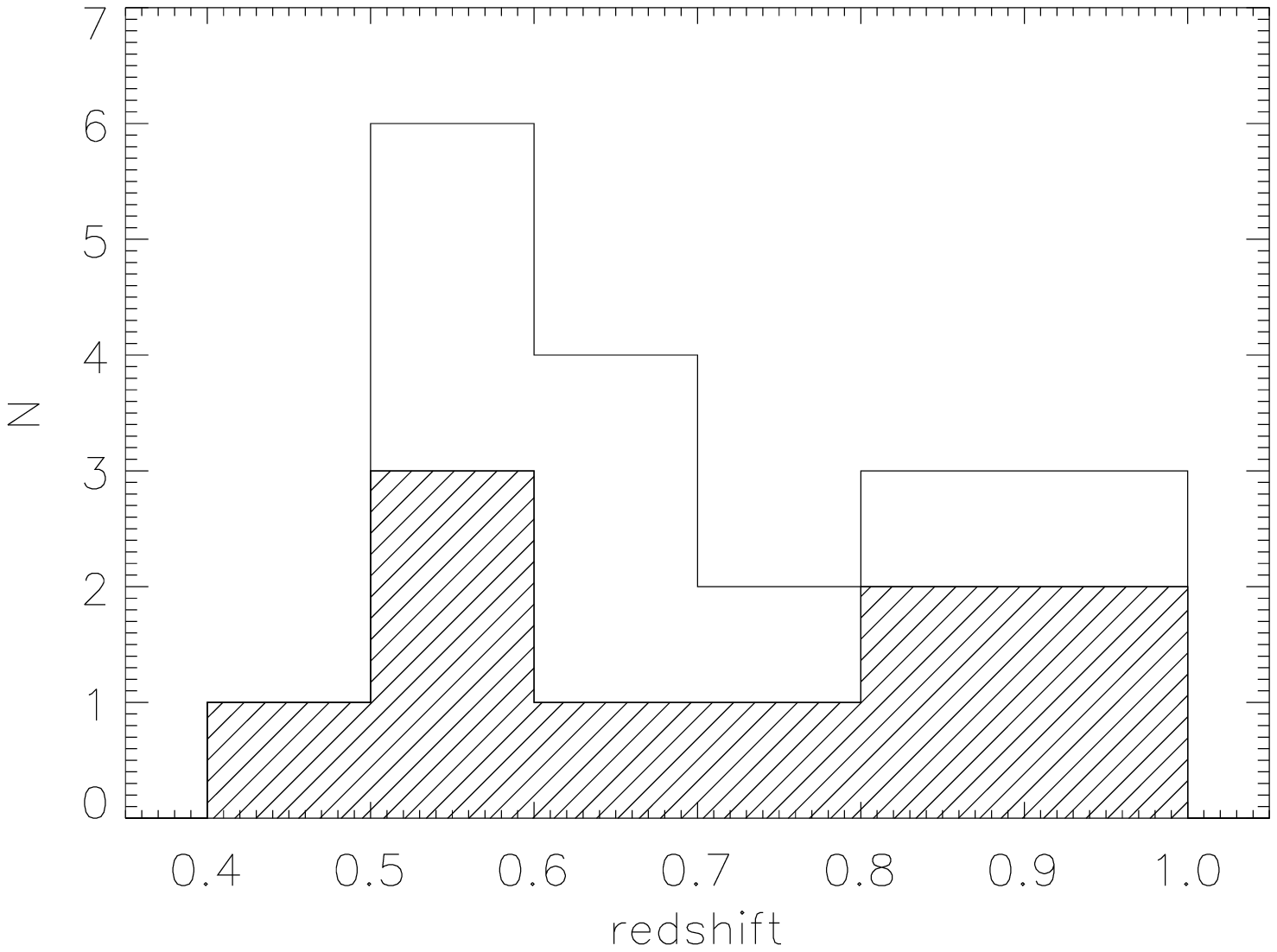}
\includegraphics[scale=0.55,angle=0,keepaspectratio]{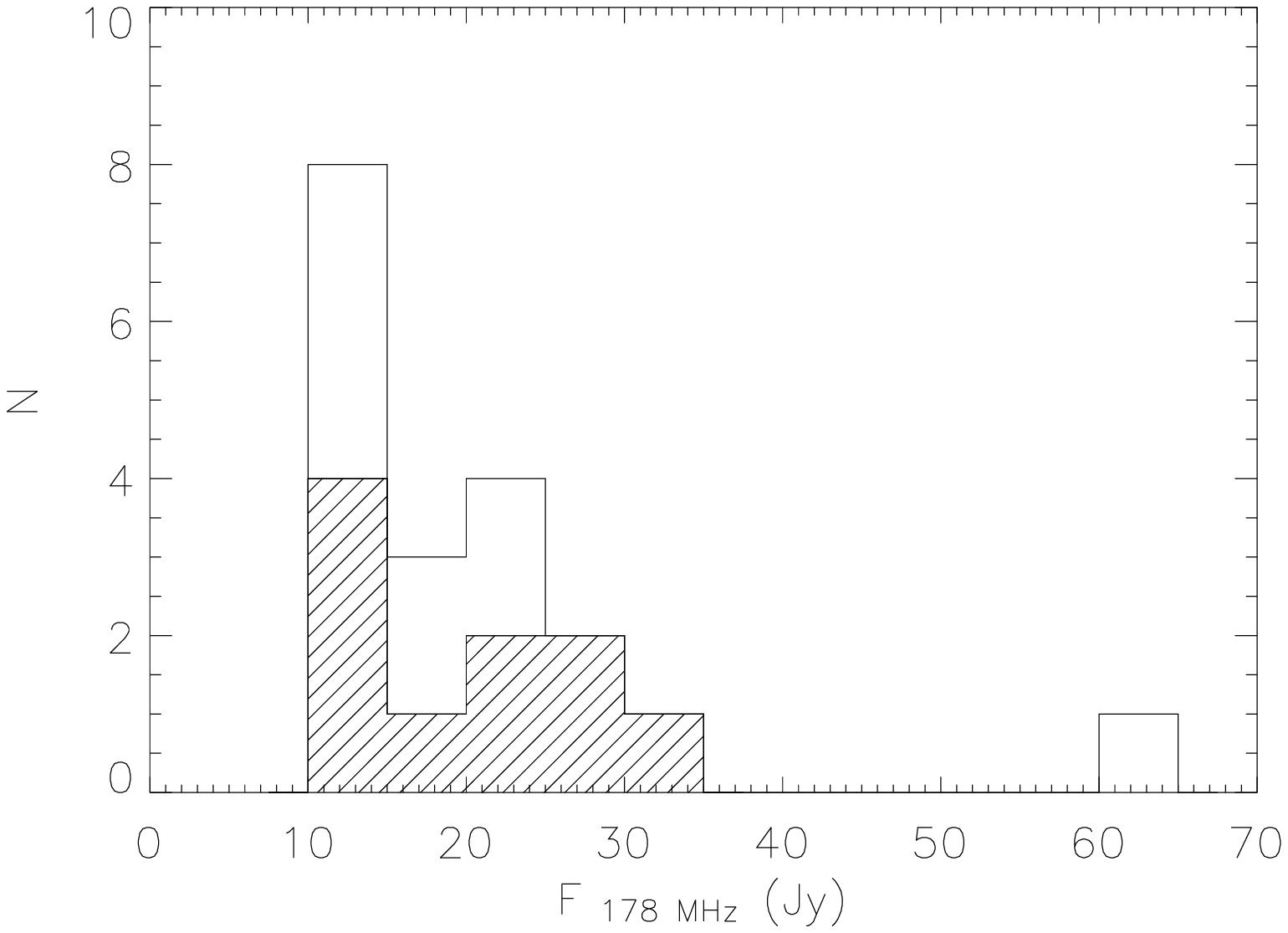}
\includegraphics[scale=0.55,angle=0,keepaspectratio]{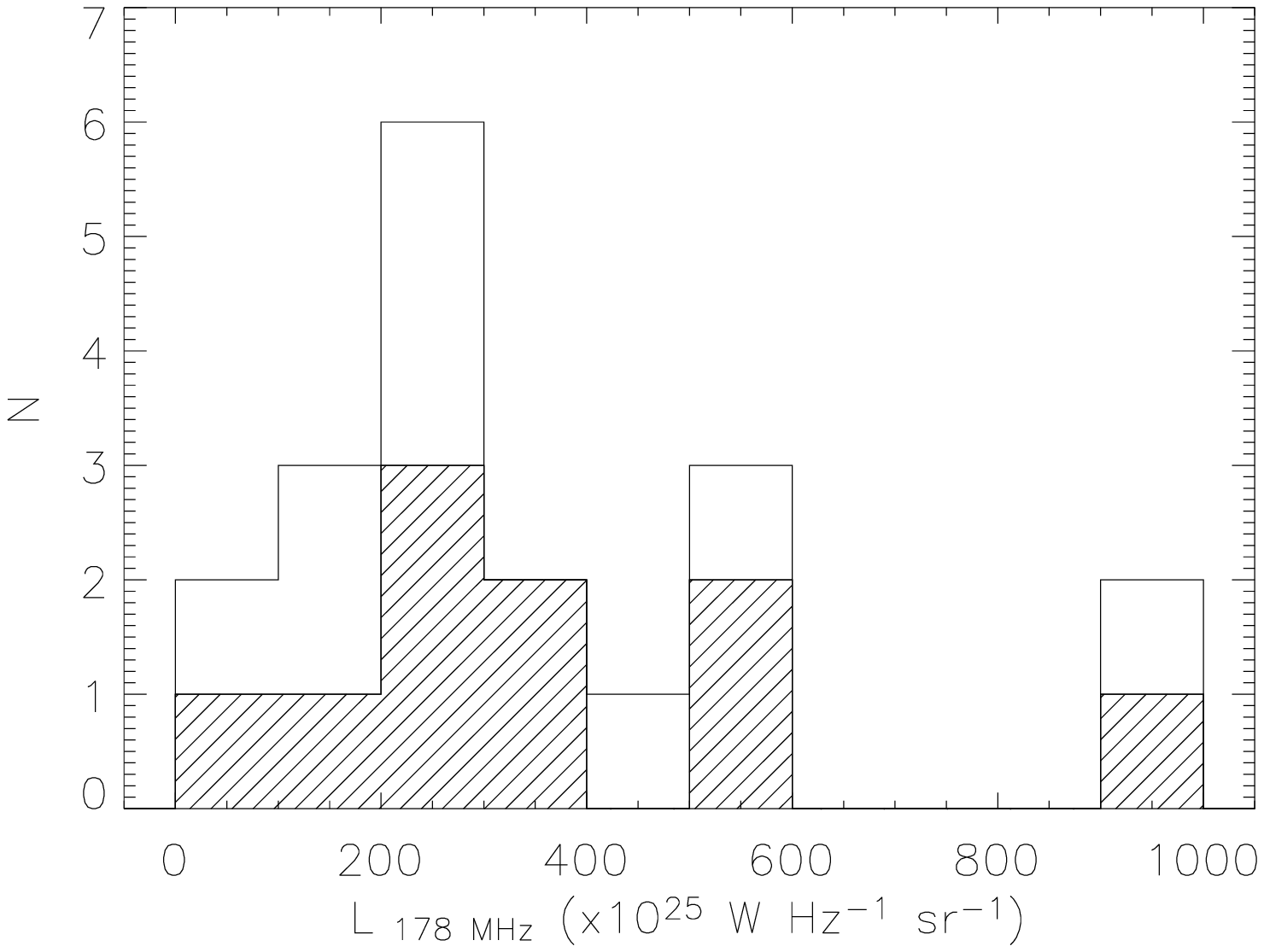}
\caption{Redshift (Top),  isotropic (178 MHz) radio flux  (Middle) and
power (bottom) for the
sample of 19 sources analysed in this paper. The shaded area
corresponds to the 10 radio galaxies.}
\label{fig:histo1}
\end{figure}

These 34 sources were selected from the 3CRR on the basis of Galactic
latitude for convenient scheduling of {\em Spitzer} observations, and
they display similar distributions in redshift and isotropic radio
power to the parent sample. 
Here we present X-ray results for 15 of the sources in this sample,
those so far observed with \cha\ or \xmm.
3C\,200 is part of the {\em Spitzer} sample although the source
lies slightly outside the $0.5<z<1$ range,  since \cha\ observations were available at
the time of the selection of the infrared sample. 
In addition, we have added to our study four  sources (3C\,228, 3C\,265, 3C\,345, 3C\,454.3) that are not in the {\em Spitzer}
sample,
since they are in the same redshift range and have available
high-quality X-ray observations. This gives us a sample of 19
  sources (see
Table~\ref{tab:sources}). Figure~\ref{fig:histo1} shows that for the
quasars and radio galaxies in  this
subsample, the redshift,
isotropic radio power, and radio flux distributions  are
similar.

\cha\ or \xmm\ observations of 9 sources (3C\,184, 3C\,200, 3C\,220.1, 3C\,228, 3C\,263,
3C\,275.1, 3C\,292, 3C\,330, 3C\,334) were awarded to us in
  support of various projects, while data for the remaining sources were extracted from the \cha\ archive.
Table \ref{tab:sources} lists the sources used in this work.

\section{X-ray observations and data preparation}\label{sec:data}

\begin{table*}
\caption{Observation log. Col 1: 3CRR name; Col. 2: Instrument used
for the observations, C stands for \cha, X for \xmm; Col.3:
Observation ID; Col. 4: Date of observation; Col 5: Detector, where
the numbers following ``ACIS'' correspond to the CCD chips turned on
during observation. The target was placed at the aim point of chip
7.  EPIC means that MOS and pn cameras have been used for the
observation and analysis; 3C\,184 was observed with MOS cameras only for
observation ID 0028540201; Col. 6: Observation mode. For
\xmm\ observations the filter is also listed and the mode refers
to the EPN camera; Col. 7: Frame
readout time; the first value for \xmm\ observations refers to MOS cameras; Col. 8: Nominal exposure Time; Col.9: Net exposure
time, after flare screening; Col. 10: Pileup fraction.}
\label{tab:obslog}
\begin{tabular}{l|crclrcccr}
\hline
Source & Instrument & Obs ID  & Obs Date &   Detector & MODE  &
TIME/FRAME & EXPOSURE & Screened time & Pileup  \\
        &       &       &       &       &       & sec   & ks & ks & fraction\\
\hline
3C\,6.1 & C & 4363 & 2002-08-26 & ACIS-235678 & VFAINT & 3.24 & 20 & 20.0& 5\%\\  
      & C & 3009 & 2002-10-15 & ACIS-235678 & VFAINT & 3.24 & 36 & 35.7& 5\%\\  
3C\,184 & C & 3226 & 2002-09-22 & ACIS-23678 &  VFAINT & 3.14 & 20 &18.9&1\%\\
      & X & 0028540201& 2001-09-19& EPIC/MOS & Thin/FF & 2.6 &38.9 & 32/--&na\\
      & X & 0028540601& 2002-03-10& EPIC & Thin/EFF & 2.6/0.2(pn)& 40.9 & 26/16.4(pn)&na\\
3C\,200 & C & 838 & 2000-10-06&  ACIS-235678& FAINT& 3.24 & 16 &14.7&1\%\\
3C\,207 & C & 2130 & 2000-11-04 &ACIS-235678& FAINT & 3.24 & 39& 37.5 &22\%\\
3C\,220.1&C & 839 & 1999-12-29 &ACIS-23678&  FAINT& 3.24 &21 & 18.5 &7\%\\
3C\,228 & C & 2453 & 2001-04-23 & ACIS-235678& FAINT & 3.24 & 12 & 10.6&2\%\\
      & C & 2095 & 2001-06-03 & ACIS-235678& FAINT & 3.24 &15.5& 13.8&3\%\\\
3C\,254 & C & 2209 & 2001-03-26 & ACIS-23678 & VFAINT& 3.14 &31& 29.5&20\%\\
3C\,263 & C & 2126 & 2000-10-28 & ACIS-235678& FAINT & 3.24 &51& 48.8 &26\%\\
3C\,265 & C & 2984 & 2002-04-25 & ACIS-235678& VFAINT & 3.24 &59&50.6 &1\%\\
3C\,275.1&C & 2096 & 2001-06-02 & ACIS-567   & FAINT  &1.64  &26& 24.8&4\%\\
3C\,280 & C & 2210 & 2001-08-27 & ACIS-235678& VFAINT & 3.24 &63.5& 46.3&$<1\%$\\
3C\,292 & X & 0147540101 &2002-10-29 & EPIC & Medium/FF & 2.6/0.07(pn)& 33.9 & 20/17(pn)&$<1\%$\\
3C\,309.1&C & 3105 & 2002-01-28 & ACIS-7&  VFAINT& 0.44 &17& 16.6&6\%\\ 
3C\,330 & C & 2127 & 2001-10-16 & ACIS-235678&  FAINT& 3.24 &44& 43.8&0\%\\
3C\,334 & C & 2097 & 2001-08-22 & ACIS-567& FAINT & 0.94 &33& 30.2&9\%\\
3C\,345 & C & 2143 & 2001-04-27 & ACIS-7  & FAINT& 0.44 &10& 9.0&13\%\\
3C\,380 & C & 3124 & 2002-05-20 & ACIS-7 & FAINT & 0.84 &5.5& 5.3&16\%\\
3C\,427.1& C & 2194 & 2002-01-27 & ACIS-235678 & FAINT & 3.24 &39&39.0 &0\%\\
3C\,454.3& C& 3127 & 2002-11-06 & ACIS-7 & FAINT & 0.84 &5.5& 5.5&24\%\\
\hline
\end{tabular}
\end{table*}

Table \ref{tab:obslog} lists basic observational information for the sources analysed in this paper.
All \cha\ observations were performed with ACIS-S with the source at
the  aimpoint of the S3 chip.  For some bright sources (see
Table~\ref{tab:obslog}) sub-array modes were used to reduce the
readout time and hence the effect of  pileup.
The \xmm\ observations for two sources, 3C\,184 and 3C\,292,  discussed in
this paper were taken with the EPIC  cameras. Data analysis  for these
sources is described in  \citet{bel04a}. 

\cha\ data were processed using CIAO 3.2.1 and CALDB v3.0.2. We first removed the {\em
afterglow} correction, and produced a new Level 1 event list. We then
generated new bad-pixel files using the CIAO tool {\sc
acis\_run\_hotpix}. The new Level 1 photon lists were then processed
using the CIAO tool {\sc acis\_process\_events} in the standard
way. The standard Charge Transfer Inefficiency (CTI) correction was not applied
for  3C\,220.1,  since the focal plane temperature was higher
than -120$^\circ$ C. For those observations telemetered in VFAINT
mode, we applied the additional background screening recommended for this
mode, although this is not particularly useful for the study of the
bright central component. We filtered for good grades (0, 2, 3, 4, 6)
and generated a Level 2 photon list after filtering using the pipeline
good-time-interval (GTI) filter.
Finally, for each source  we produced a light curve in a region void of sources to check for variability of the background. No strong flare events were found in any of our sources. However, all data sets were filtered following the prescription given in {\tt http://asc.harvard.edu/ciao/threads/filter/}.
We then applied an astrometry correction to the final event list using
the tool available on
line\footnote{http://asc.harvard.edu/cal/ASPECT/fix\_offset/fix \_offset.cgi}.
The level 2 event lists were then used for the analysis described below.

\section{Analysis}
\subsection{Spatial analysis}
We performed preliminary spatial analysis on the  \cha\ data in order to investigate whether the X-ray
radial profile was consistent with the instrumental point spread
function (PSF). PSFs for each source were generated using the PSF
library for ACIS at a convenient  energy depending on the spectrum
of the source. For most sources this was 2.5 keV. We then used the
CIAO tool {\sc mkpsf} to simulate a point source image, and the tool
{\sc dmimgcalc} to normalize the PSF to the count rate of each
source. We extracted radial profiles for each source in the
energy range 0.5-8.0 keV and compared with the radial profile derived
for the simulated PSF. Features such as jet emission or
extended emission clearly associated with the radio lobes were masked
prior to the radial profile extraction. We are aware that this
approach is not perfect 
since a monochromatic estimate of the PSF was used. However,
it gives an estimate of the extent of the PSF that is good
  enough to allow us to define the radius for spectral
analysis of the nuclear region and the associated background for each
source, especially in cases where the sources are piled up.

This quick spatial analysis also gives indications of the presence of
extended emission 
associated with a cluster-like environment or inverse-Compton
scattering in the radio lobes \citep{croston05}. We will discuss the
environments of these sources in a future paper.

We found that for most of the sources more than 90 per cent of the
flux was included in a circular region of radius 2.5 pixels. 
The radial profile of 3C\,200 
shows that extended emission starts to dominate at $\sim 2$ arcsec from
the X-ray peak. However, only 2 counts contribute
to the extended flux in the circular region used for the spectral
analysis, and this is much less than any statistical
error.
For 3C\,263 we found that the point source photon distribution 
was broader than the PSF obtained at 2.5 keV, which is
  likely to be the effect of the strong pileup (see Sect. \ref{sec:spfit}): a piled-up point
source will show a photon distribution which is  flatter in the centre
and broader in the wings than a standard  PSF as calculated with {\sc mkpsf}. 
For all sources, we applied
 corrections for the missing fraction of the PSF (see below) to
 the X-ray fluxes we quote.

\subsection{Timing analysis}

For all \cha\ observations we generated light curves from the 0.5-8.0 keV energy range for each
core using the CIAO tool {\sc dmextract}, and in different time bins
from 100 to 4000 s intervals, depending on the source exposure
time and photon statistics (we did not perform timing analysis for the
\xmm\ observation of 3C\,292).

\begin{table}
\begin{center}
\caption{X-ray variability}\label{tab:timing}
\begin{tabular}{l|crc}
\hline
Source & time bin  & $\chi^2_r$/d.o.f. & $P_{\chi^2}$ \\
        & s     &       &       \\
\hline
3C\,6.1/3009 & 750&0.599/27 &0.950\\  
3C\,6.1/4363 & 2000&0.663/18 &0.850\\  
3C\,184 & 3000&0.450/6 &        0.846 \\
3C\,200 &       2000&0.650/7 &0.715\\
3C\,207 & 1000 &  0.521/38&0.993\\
3C\,220.1&650&0.806/30 & 0.764 \\
3C\,228/2095 & 3000& 1.012/5&0.962\\
3C\,228/2453 & 2000& 0.600/5&0.700 \\
3C\,254 & 3000&0.568/10&0.842\\
3C\,263 &       500&0.470/100&1.000\\
3C\,265 & 3000&0.648/17&0.856\\
3C\,275.1& 750&38.549/10&0.000\\
3C\,280 &       3000&0.500/14&0.935  \\
3C\,309.1& 3000&0.612/6 &0.721\\
3C\,330 & 3000&0.464/15&        0.959\\
3C\,334 & 2000&0.496/17 &0.956\\
3C\,345 & 750&13.305/14&0.503\\
3C\,380 & 650   &1.286/9&0.238\\
3C\,427.1& 4000&0.300/9 &0.975\\
3C\,454.3&1000&0.711/5  &0.615\\
\hline
\end{tabular}
\vskip 8pt
\begin{minipage}{6cm}
Numbers given after multiple observations of the same source are the
{\it Chandra} observation IDs. The time bin is the one that gives the
largest $\chi^2$ value (see the text). The probability $P_{\chi^2}$ is
the probability of obtaining a value of $\chi^2$ as extreme as this
under the null hypothesis of no variability.
\end{minipage}
\end{center}
\end{table}

 We used $\chi^2$ to test for significant deviations from the
average count rate within the light curve, and  Table \ref{tab:timing}
reports these $\chi^2$ values calculated using
the binning that shows  the largest deviation from a constant value as
well as the probability that the light curve is constant.
None of our sources exhibits significant time variability, including
3C\,207 \citep{hough02}, and the core-dominated quasars which are known to be variable at 5 GHz (and
higher frequency)  on time scales of
years (e.g.,
\citealt*{ltp01,aah03}). 3C345 in particular was found to be variable
in X-ray flux density \citep{unwinetal97}, but on timescales of order
of weeks. 

We investigated flux variability for the two sources which were
observed in two exposures, 3C\,6.1 and 3C\,228. The average count rate
for the two exposures  of 3C\,6.1 are 0.044$\pm0.008$  s$^{-1}$ and
0.039$\pm0.014$ s$^{-1}$ for exposure IDs 3009 and 4363,
respectively.
3C\,228 does not show any flux variability either, with both exposures
having count rate of 0.023$\pm0.005$ s$^{-1}$. 

\subsection{X-ray spectral fitting}\label{sec:spfit}
The spectrum of each core was  extracted from a circle of radius
2.5 pixels (1.23 arcsec; core region), which includes $>90$ per cent of the
ACIS-S3 PSF (as verified by our spatial analysis). A local background
region was selected from an annulus of internal and external radii 3.5
and 6 pixels.  This choice of background should provide the
best possible subtraction of 
any contribution from a cluster-like environment. In all cases the
core strongly dominates any such extended emission. The net counts in the core region in the 0.2-10.0 keV energy band, for each source, are listed in Table \ref{tab:corehard}. 

Spectra were extracted  using the tool {\sc psextract}, which
generates Ancillary Response Files (ARF) and Redistribution Matrix
Files (RMF). Spectra were grouped to have a minimum of 15 to 20 counts
per channel (except for the \cha\ observations of 3C\,184 and 3C\,427.1
where there are too few counts for spectral fitting). We used version  3.0.2 of the calibration database, which accounts for spatial variation of the  Quantum Efficiency (QE) degradation.

Because of the nature of the sources discussed here, the issue of
  pileup needed to be considered. Pileup  
will distort the
spectrum at high energy (for details see \citealt{ballet99};
\citealt{davis01pileup}). An indication of the pileup fraction (PUF;
defined as the ratio of the number of detected events that consist of
more than one photon to the total number of detected events) was
estimated using PIMMS and Figure 6.18 in the \cha\ Proposers'
Observatory Guide
Sect. 6.14\footnote{http://asc.harvard.edu/proposer/POG/html/ACIS.html\\
\#SECTION037140000000000000000}. The PUF was calculated for
each source, and is listed in Table \ref{tab:obslog}. Pileup effects were ignored if the PUF was less than
8 per cent, although we verified that the addition of a pileup model in
{\sc XSPEC} did not change the best-fit parameters for such sources (see
Sect. \ref{sec:resultsfit} for more details). For those sources for which a higher PUF was found, we excluded a
circle of radius from 1 to 2 pixels (see below for details of each
individual source) from the initial circle of radius 2.5 pixels and used this
annulus as the region
for spectral analysis, since the wings of the PSF are less affected by
pileup. Although an annular region contains only a fraction of the
PSF, it is valuable in finding the correct spectral index of the
source. To measure fluxes and flux densities for these sources we
adjusted  the normalization to account for missing counts based on
simulating the PSF at
the position of the source using MARX. Although this
correction is dependent on the source spectrum, it has been shown that
for NGC 6251 the spectral dependence is negligible
\citep{devansngc6251}. We applied similar tests to those applied for
NCG\,6251 to  3C\,334, and found that
any correction due to the energy-dependence of the PSF is within the
statistical errors. Therefore we ignored energy dependence  when
correcting fluxes and flux densities for the missing PSF, and we
corrected solely for the fraction of the PSF missing at the
energy of 2.5 keV. 

In the case of 3C\,263, which is heavily piled-up, we  adopted
the spectrum extracted in an annulus of inner and outer
radii 3 and 5 pixels, after inspecting the radial profile. 

Spectral analysis was performed using {\sc XSPEC} v11.3.1. All spectra
were fitted in the energy band 0.5-8.0 keV, where the ACIS-S
calibration is most precise. We initially fitted the
background-subtracted core spectrum with a model composed of Galactic
absorption (see Table \ref{tab:sources}), and a power law. 
Additional components were added to the model as required after inspection of the
residuals. The significance of adding additional model components to
the fit was estimated using the $F$-test.

Fluxes, flux densities and luminosities for all sources  whose
spectrum was extracted in the canonical circle of radius 2.5 pixels
were corrected by 10 per cent to
account for the missing PSF fraction. Values derived from spectra extracted in annular regions were
corrected appropriately as described above. Moreover, as a conservative approach, we added a 10 per cent
systematic error  to the error estimates for the fluxes, flux
densities and luminosities of all sources with  pileup fraction
greater than 8 per cent (see Table~\ref{tab:obslog}).
\section{Results}
\begin{figure*}
\includegraphics[scale=0.7,angle=0,keepaspectratio]{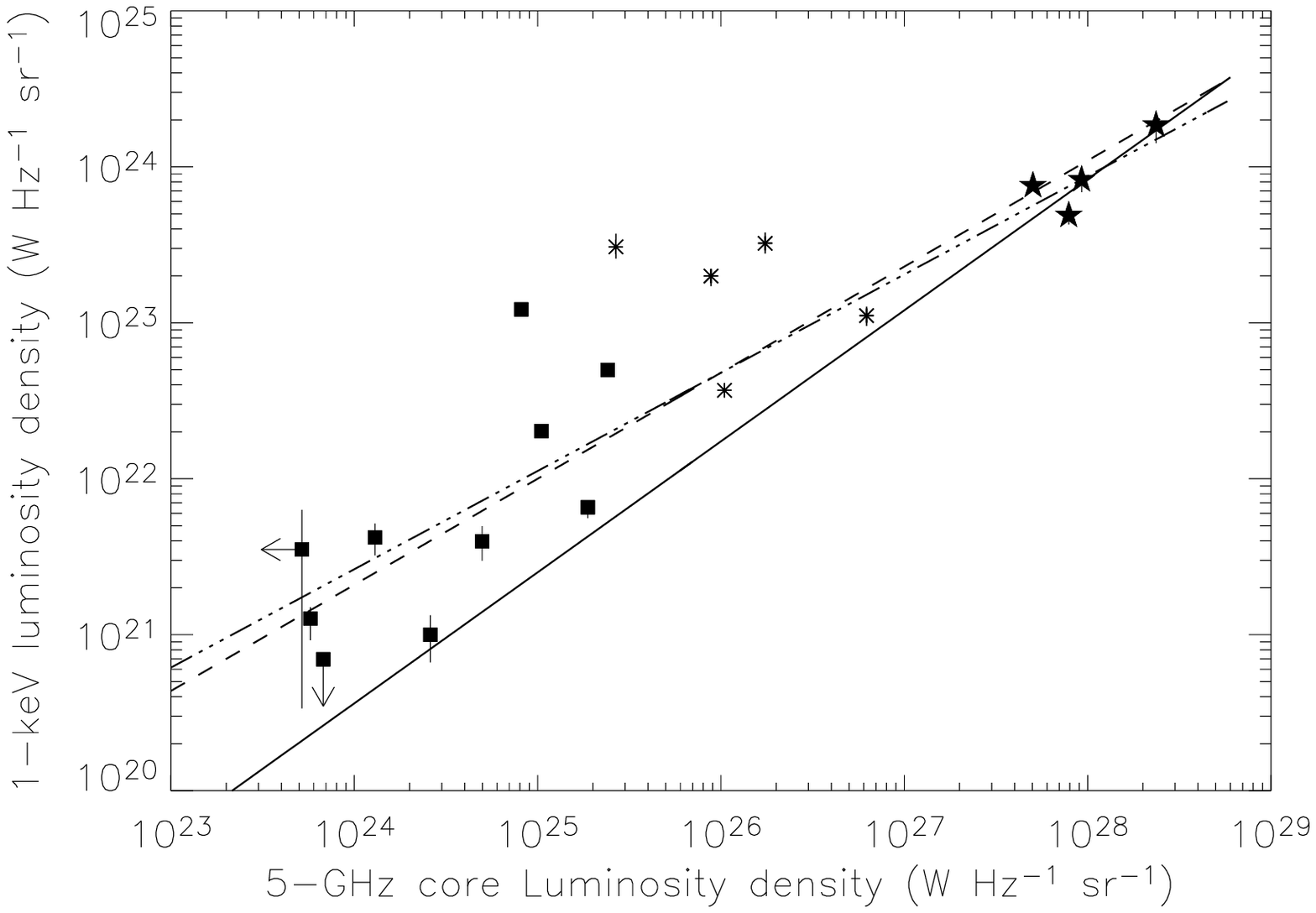}
\caption{
  1-keV X-ray core versus 5-GHz radio core spectral luminosity density. Symbols are
as follows: filled squares are radio galaxies (when applicable only the unabsorbed component
was taken into account to compute the 1-keV luminosity density); asterisks
are lobe-dominated quasars and stars are core-dominated quasars. The
dot-dashed line corresponds to the maximum likelihood linear regression
method applied to the whole sample. The dashed line is the
 linear regression applied to the sample composed of
core-dominated quasars and radio galaxies only. The continuous line
corresponds to the best fit regression for core-dominated quasars
only. When 1$\sigma$ error
bars are not visible they are smaller than the symbols. }
\label{fig:2fluxcorr}
\end{figure*}

\begin{table*}
\caption{Spectral fit results for the core region. {\bf Col.~1:} source 3CRR
name; {\bf Col.~2:}  net counts in the 0.2--10.0 keV energy band contained
in a circular region of radius 2.5 pixels; if two exposures have
been used this is the sum of the two data sets; {\bf Col.~3:} intrinsic \nh,
when required; {\bf Col.~4:} power law index. This corresponds to the slope
of the hard spectrum when a composite model is used; {\bf Col.~5:}
normalization of the power law model, in units of photons keV$^{-1}$
cm$^{-2}$ s$^{-1}$ at 1 keV. The normalization corresponds to the
best-fit value found while fitting the region specified in Col. 9 and
does not correspond to the final luminosity displayed in
Col. 8. Moreover, the normalization
corresponds to the power law used to fit the hard spectrum when a
model composed of more than one power-law is adopted; {\bf Col.~6:}
$\chi^2$/d.o.f. corresponding to the best-fitting model;
{\bf Col. 7:} where the best-fitting model was not a power law with
  Galactic absorption, this column gives the probability under the
  null hypothesis (from the $F$-test) of obtaining an improvement in
  $\chi^2$ as large as was observed. When not applicable the
original model represents a good fit to the data; {\bf Col.~8:} unabsorbed
X-ray Luminosity in the energy band 2.0-10.0 keV - absorbed component
only in radio galaxies (flux for the same
component is listed in Table \ref{tab:1kevflux}); {\bf Col.~9:} comments; if
there are no comments, the region used for spectral analysis is a
circle of radius 2.5 pixels.}\label{tab:corehard}
\begin{tabular}{l|rcclccrl}
\hline
Source  & net counts    & \nh (intrinsic) & Gamma &  Norm & $\chi^2$/d.o.f. & $F$-test  &  $L_{\rm X}$[2-10 keV] & Comments\\
        & [0.2-10 keV]  & $\times10^{22}$ cm$^{-2}$ &  &  $\times 10^{-5}$&  & &$\times10^{44}$ &\\
        &               &                 & & (ph/keV/cm/s)& & &(~erg s$^{-1}$)& \\
\hline
3C\,6.1 & 2326$\pm49$ & 0.28$^{+0.11}_{-0.10}$&1.53$^{+0.06}_{-0.07}$&6.5$^{+0.4}_{-0.5}$ &  139.1/126 &3.6$\times10^{-3}$& 10.0$^{+0.7}_{-0.6}$ & 2 exposures combined \\  
3C\,184$^a$ & 776$\pm65$&  48.7$^{+22.0}_{-12.1}$ &1.4$^{+0.3}_{-0.2}$&2.4$^{+1.1}_{-1.0}$ & 39.3/44&2.7$\times10^{-7}$&5.7$^{+2.2}_{-2.0}$ & XMM  \\
3C\,200 & 166$\pm14$  & na  & 1.68$^{+0.18}_{-0.16}$ &  1.3$^{+0.1}_{-0.2}$ &  5.3/5 & na &0.41$^{+0.07}_{-0.08}$ & \\
3C\,207 & 6349$\pm80$ & na& 1.27$^{+0.05}_{-0.05}$ &3.6$^{+0.1}_{-0.1}$&  96.5/105 & na  &13.7$^{+1.7}_{-2.1}$& 1.0-2.5 pix annulus\\
3C\,220.1& 1133$\pm34$& na & 1.55$^{+0.04}_{-0.05}$ & 5.4$^{+0.2}_{-0.2}$ &55.3/57 & na & 3.9$^{+0.2}_{-0.1}$& \\
3C\,228 &  646$\pm50$  & na &1.59$^{+0.11}_{-0.10}$&2.7$^{+0.2}_{-0.2}$ &  23.7/32 & na &1.5$^{+0.1}_{-0.1}$ & 2 exposures combined \\
3C\,254 & 4528$\pm70$ & na & 1.64$^{+0.11}_{-0.10}$ &1.4$^{+0.2}_{-0.1}$&  34.1/33 & na& 20.8$^{+4.4}_{-3.1}$ & 1.5-2.5 pix annulus\\
3C\,263 & 9759$\pm100$ & na& 1.88$^{+0.10}_{-0.10}$& 1.10$^{+0.08}_{-0.04}$&  14.2/20 &  na& 15.0$^{+2.5}_{-1.9}$ & 2.5-5.0 pix annulus \\
3C\,265 & 285$\pm17$ & 16.8$^{+10.5}_{-7.2}$ &1.01$^{+0.66}_{-0.30}$  & 1.3$^{+2.5}_{-0.8}$ &  5.6/10 &  3.2$\times10^{-3}$&3.1$^{+1.2}_{-1.2}$& \\
3C\,275.1& 1302$\pm36$& na& 1.45$^{+0.05}_{-0.05}$ & 5.1$^{+0.2}_{-0.2}$ &  62.0/63 & na &3.3$^{+0.2}_{-0.1}$&  \\
3C\,280$^c$ &   53$\pm8$ & 9.7$^{+6.2}_{-4.7}$ & $1.27^{+0.43}_{-0.41}$  &$0.4^{+0.1}_{-0.1}$  & 0.7/3  & 2.0$\times10^{-2}$ &10.0$^{+3.5}_{-3.5}$& \\
3C\,292$^b$ &  251$\pm10$& 26.4$^{+12.8}_{-5.9}$ & 2.05$^{+0.15}_{-0.15}$& 4.5 $^{+1.8}_{-1.3}$&14.6/19 &1.8$\times10^{-4}$ &2.5$^{+0.7}_{-0.7}$ & XMM \\
3C\,309.1& 5528$\pm75$ & na & 1.54$^{+0.02}_{-0.03}$ &  34.2$^{+0.5}_{-0.7}$ & 159.4/169 & na & 60.0$^{+2.0}_{-2.0}$ & \\
3C\,330& 118$\pm11$  & 23.6$^{+16.3}_{-13.4}$& 1.95$^{+0.55}_{-0.36}$& 2.3$^{+1.4}_{-1.8}$ &6.7/6& 1.3$\times10^{-1}$& 0.8$^{+0.2}_{-0.6}$ & \\
3C\,334 & 7241$\pm85$ & na &1.74$^{+0.04}_{-0.05}$&8.2$^{+0.2}_{-0.3}$&105.4/78 & na&12.0$^{+2.2}_{-2.2}$ & 1.0-2.5 pix annulus\\
3C\,345 & 6637$\pm81$& na&1.27$^{+0.11}_{-0.11}$ &15.8$^{+2.3}_{-2.1}$&87.8/83&1.2$\times10^{-2}$ &43.3$^{+7.7}_{-5.0}$ & 1.0-2.5 pix annulus\\
3C\,380 & 2553$\pm51$ &na&1.54$^{+0.09}_{-0.09}$&12.5$^{+1.2}_{-1.0}$&15.5/20&na&65.0$^{+12.5}_{-12.5}$& 1.5-2.5 pix annulus \\
3C\,427.1 &15$\pm3$ &10 (fix) &1.6 (fix)& 0.2$^{+0.1}_{-0.1}$  & 2.7$^d$  &na& 0.13$^{+0.05}_{-0.05}$ &  $C$-statistics used \\
3C\,454.3$^f$& 3889$\pm63$  & na & 1.25$^{+0.15}_{-0.15}$ &5.3$^{+0.7}_{-0.7}$ &  9.0/8 & na &235.0$^{+60.0}_{-60.0}$& 2.0-2.5 pix annulus\\
\hline
\end{tabular}
\vskip 8pt
\begin{minipage}{500pt}
\cha\ data fits were done in the 0.5-8.0 keV band; ``2 exposures'' in Col. 9 means
that the spectra from both exposures of the source have been
used. Luminosities are corrected for the extraction aperture while normalizations are
the actual, best-fit value. $^a$ Parameters are from \xmm\ data fitting. The total spectrum of the source was extracted in a
circle of radius 40 arcsec; the count rate in Col. 2 corresponds to the
total EPIC count rate in this region (see \citealt{bel04a} for
details); the \cha\ observation of 3C\,184 contains 41 counts in the
energy band [0.2-10.0] keV. The \xmm\ and \cha\ fitting results
are in agreement, but the former was preferred since gives
significantly better constraints. $^b$ The \xmm\ core spectrum was extracted in a circular region
of radius 10 arcsec, and the count rate is the total EPIC
background-subtracted count rate in this
region (the background estimated in an annulus of inner
radius 30 arcsec and outer radius 35 arcsec -- see \citealt{bel04a} for details); the ARF accounts for the missing PSF, and the spectrum
was fitted between 0.3 and 10 keV. $^c$ An $F$-test was performed for a model with intrinsic absorption versus a model with no intrinsic
absorption; see the details for this source in Appendix A for discussion of an
unabsorbed component at soft energies. $^d$ $C$-statistic value
instead of $\chi^2$.  The quoted error is the
``real'' error, i.e. statistical plus systematic.
\end{minipage}
\end{table*}

\subsection{X-ray spectral properties}\label{sec:resultsfit}

\begin{table*}
\begin{minipage}{340pt}
\begin{center}
\caption{Spectral fit results for the core region: Additional components}\label{tab:coresoft}
\begin{tabular}{l|cc|cc|cr}
\hline
Source  &\multicolumn{2}{c|}{Soft PL} &  \multicolumn{2}{c|}{Thermal}&   Flux   & \multicolumn{1}{c}{$L_{\rm X}$}       \\
\cline{2-7}
        & $\Gamma$ & $N_{PL}$ &k$T$ & $N_{Th}$  & \multicolumn{2}{c} {[0.2-10.0] keV}   \\
        &       &$\times10^{-5}$        & (keV) & $\times10^{-5}$       &(erg~s$^{-1}$~cm$^{-2})$       & (erg~s$^{-1}$) \\
\hline
3C\,184$^a$ & 1.9$^{+1.4}_{-1.5}$ & 0.1$^{+0.2}_{-0.09}$ & 3.6$^{+14.1}_{-1.8}$&2.3$^{+0.5}_{-0.5}$& $7.3\times10^{-15}$& $3.0\times10^{43}$  \\
3C\,265 & 2.1$^{+0.8}_{-0.7}$ & 0.17$^{+0.04}_{-0.04}$ &  --- &  --- &1.0$\times10^{-14}$ & 2.8$\times10^{43}$\\
3C\,280 & 2.2(fix) & 0.02$^{+0.01}_{-0.01}$ &  --- & --- & $1.5\times10^{-15}$ & 7.1$\times10^{42}$\\
3C\,292 & 2.7$^{+0.6}_{-0.7}$ & 0.2$^{+0.1}_{-0.1}$& --- & ---  & 1.3$\times10^{-14}$&4.2 $\times10^{43}$  \\
3C\,330 & 1.9$^{+0.8}_{-0.7}$& 0.16$^{+0.03}_{-0.03}$ & --- & --- &1.0$\times10^{-14}$ & 1.2$\times10^{43}$\\
3C\,345 &1.8$^{+0.1}_{-0.1}$ &21.3$^{+0.6}_{-0.7}$ & ---& --- &  5.0$\times10^{-12}$&6.0$\times10^{45}$ \\
3C\,427.1& 2.2(fix) & $<0.08$ & ---& ---& $<4.5\times10^{-15}$ &$<5.6\times10^{42}$ \\
\hline
\end{tabular}
\end{center}

Upper limits are quoted at the $3\sigma$ confidence level for one interesting
parameter; errors are the 1$\sigma$ confidence range. $^a$ Flux and luminosity 
refer to the soft power-law component. The soft power-law parameters were obtained by fitting the \cha\ spectrum. The
bolometric X-ray luminosity of the thermal component is
8.3$\times10^{43}$ erg s$^{-1}$ \citep{bel04a}.
\end{minipage}
\end{table*}

The results of the spectral fitting are reported in
Table~\ref{tab:corehard}. A single power law absorbed by Galactic
absorption provides a good fit to the core continuum for 11 sources
(3C\,200, 3C\,207, 3C\,220.1, 3C\,228, 3C\,254, 3C\,263, 3C\,275.1, 3C\,309.1, 3C\,334, 3C\,380,
3C\,434.3), of which 8 are lobe-dominated  or core-dominated quasars.  
Intrinsic absorption is required for 7 sources, all of them radio
galaxies (3C\,6.1, 3C\,184, 3C\,265, 3C\,280, 3C\,292, 3C\,330, 3C\,427.1). An additional
thermal component is required for only 1 object, 3C\,184,
a very small radio source observed with \xmm\ for which  cluster-scale X-ray emission was detected
\citep{bel04a}.  This is much less than any statistical error.
In only one case, 3C\,345, is a single power law not an acceptable fit
for a quasar. A second power law is necessary to account for the
flattening above 1.7 keV in the core spectrum. This is in
agreement with the results of \citet{gambill03}, who used a broken power
law to fit the spectrum. Several of the radio galaxies required a soft
unabsorbed power law in addition to the absorbed power law. Details of
additional continuum components, both
non-thermal and thermal, required to
fit the core spectra of the sources in our sample are listed in Table~\ref{tab:coresoft}.

In agreement with \cite{bondi04}, we detect significant line emission for  3C\,265, indicating a
neutral Fe line at a  rest-frame energy of 6.37$\pm0.13$ keV  and of equivalent width
443$^{+196}_{-190}$ eV. The core spectrum of 3C\,330 displays a broad, asymmetric line at
rest-frame energy 6.7$^{+0.6}_{-0.3}$ keV. Although this may suggest the presence of ionized iron in the central
region, deeper observations of this source are needed to establish the
reality of the line.
Hints of
possible line emission are seen in  3C\,207 (ionized, see also
\citealt{brunetti02}) and 3C\,275.1 (neutral),
although the addition of a line does not technically improve these fits.

\begin{table*}
\centering
\caption{X-ray and radio fluxes and flux densities}\label{tab:1kevflux}
\begin{minipage}{460pt}
\begin{tabular}{l|ccccccc}
\hline
Source  & $f_{\rm 1 keV}$& $L_{\rm 1 keV}\times10^{22}$ &$f_{\rm 2.0-10.0 keV}$& core $f_{\rm 5 GHz}$   & core $L_{\rm 5 GHz}\times10^{24}$ & total $L_{\rm 178 MHz}\times10^{25}$ & $\alpha_R$\\
        & (nJy)& (W Hz$^{-1}$ sr$^{-1}$) & ($\times10^{-13}$ erg cm$^{-2}$ s$^{-1}$) & (mJy) &
(W Hz$^{-1}$ sr$^{-1}$) & (W Hz$^{-1}$ sr$^{-1}$) &\\

\hline
3C\,6.1 & 47.6$^{+2.9}_{-3.7}$ & 12.18$^{+0.74}_{-0.95}$&3.9$^{+0.5}_{-0.4}$& 4.4 & 8.2 &332 &0.68\\  
3C\,184 &$0.73^{+0.58}_{-0.66}$ & 0.35$^{+0.28}_{-0.32}$&1.7$^{+0.7}_{-0.6}$& $<$0.2 &$<$0.5 &535 &0.86\\
3C\,200 & 9.5$^{+0.8}_{-1.4}$ & 0.66$^{+0.06}_{-0.1}$ &0.6$^{+0.1}_{-0.1}$&35.1&18.8&71.7 &0.84\\
3C\,207 & 79.1$\pm11$ & 11.11$^{+1.55}_{-1.55}$ &9.7$^{+1.3}_{-1.3}$ &510$^V$ &622.5&229 &0.90\\
3C\,220.1&39.8$^{+1.3}_{-1.4}$ & 4.99$^{+0.16}_{-0.18}$ &2.7$^{+0.1}_{-0.1}$& 25 &24.1&210 &0.93\\
3C\,228 &19.8$^{+2.0}_{-1.3}$& 2.02$^{+0.20}_{-0.13}$ & 1.4$^{+0.1}_{-0.1}$&13.3 & 10.5&231 &1.0\\
3C\,254 &153.3$^{+32.9}_{-24.1}$ &30.67$^{+6.58}_{-4.82}$ &10.4$^{+2.2}_{-1.4}$&19 &26.7&412 &0.96\\
3C\,263 & 187.2$^{+32.0}_{-26.1}$ &32.31$^{+5.52}_{-4.51}$ &9.0$^{+1.2}_{-1.4}$&157 & 174.2&220&0.82\\
3C\,265$^a$ & 1.2$^{+0.3}_{-0.3}$&  0.40$^{+0.09}_{-0.10}$   &1.8$^{+0.3}_{-0.5}$  &2.9 &5.0&515&0.96\\
3C\,275.1&37.6$^{+1.4}_{-1.2}$&3.69$^{+0.14}_{-0.12}$ &3.4$^{+0.1}_{-0.1}$ &130$^V$ &104.5&194&0.96\\
3C\,280$^a$ & 0.18$^{+0.08}_{-0.09}$ & 0.10$^{+0.03}_{-0.03}$&$0.3^{+0.1}_{-0.1}$ &1.0(7\%) &2.6&931&0.81\\
3C\,292$^a$ & 1.3$^{+0.3}_{-0.3}$ &0.42$^{+0.09}_{-0.1}$  &1.1$^{+0.2}_{-0.2}$ &1.0(11\%) &1.3&180&0.80\\
3C\,309.1&250.4$^{+4.0}_{-5.3}$ &75.90$^{+1.21}_{-1.60}$ &19.9$^{+0.2}_{-0.3}$ &2350 & 5030.0&591&0.53\\
3C\,330 &1.1$^{+0.2}_{-0.2}$ &0.13$^{+0.02}_{-0.03}$ &0.7$^{+0.2}_{-0.5}$ & 0.74(6\%)& 0.6&255&0.71\\
3C\,334 &181.0$^{+19.0}_{-25.0}$&19.94$^{+2.09}_{-2.75}$&10.0$^{+2.0}_{-2.0}$ &111&88.2&110&0.86\\
3C\,345 &470.4$^{+62.5}_{-60.3}$& 48.89$^{+6.49}_{-6.27}$&46.0$^{+7.6}_{-13.6}$&8610$^V$&7889.9&97.2&0.27\\
3C\,380 &500.3$^{+80.3}_{-84.5}$ &82.72.0$^{+13.28}_{-13.97}$&40.0$^{+8.0}_{-8.0}$ & 7447 &9272.1&930&0.71\\
3C\,427.1 & $<0.5$ & $<0.07$ &0.1$^{+0.04}_{-0.04}$ &0.8(4\%) &0.68&302&0.97\\
3C\,454.3&  820$^{+192}_{-192}$ &185.49$^{+43.43}_{-23.63}$ &100$^{+21.0}_{-14.0}$ &12200$^V$&2.36$\times10^4$&224&0.04\\
\hline
\end{tabular}

$^a$ Flux density at 1 keV is computed by using the soft component only, upper
limits when appropriate. The
2.0-10.0 keV flux is from the absorbed component only in the case of
radio galaxies. $^V$ the source is
known to be variable at the measured frequency. Errors are quoted
at 1$\sigma$ confidence for one interesting parameter; upper limits are
3$\sigma$ confidence level. Unless otherwise stated, errors on the 5 GHz flux densities are
3\%, as defined by calibration errors.
\end{minipage}
\end{table*}

\subsection{X-ray versus radio correlations}\label{sec:fluxflux}

In Table \ref{tab:1kevflux} we list the unabsorbed X-ray flux density at 1 keV,
derived from our spectral fitting, and the 5-GHz core flux density,
obtained from various sources and tabulated in the on-line 3CRR
catalogue\footnote{http://www.3crr.dyndns.org/cgi/database}.
Errors on the radio measurements are generally dominated by the
calibration errors (of order 3 per cent)
and are thus small compared to the X-ray errors. Where the
  uncertainty is dominated by thermal noise we tabulate per-source
  uncertainties in Table~\ref{tab:1kevflux}. We have also computed the  1-keV and 5-GHz rest frame spectral
luminosity densities, assuming $\alpha_X = \Gamma-1$ where $\Gamma$ is
the best-fit spectral index. In general the spectral indices of
the radio cores are not known (and likely to be variable) but
as compact cores are known to have typical spectral indices close to 0
\citep[e.g.][]{bf78} we adopt $\alpha_R = 0$ in calculating the
rest-frame radio luminosities.
We used the analysis package {\sc asurv} Rev 1.1
\citep{asurvref}, to calculate the generalized Kendall's $\tau$
coefficient in the presence of censored data to the relation between core 1-keV
flux density and the core 5~GHz flux density. We found that the
flux-flux correlation is significant
for the sample as a whole, as well as for subsamples of data. This
support the idea that the luminosity-luminosity correlation is not redshift-induced.
However, to test correlations between properties of sources spanning a wide
redshift range, and particularly for high-redshift sources for which
$K$-correction must be taken into account, the respective luminosity
values should be used (e.g. \citealt*{kemb86}).
 We apply the generalized Kendall's $\tau$ algorithm to the
1-keV luminosity density and 5-GHz luminosity density correlation (see Table~\ref{tab:2lumcorr}).
Linear regression  were evaluated by applying a maximum-likelihood fitting routine to fit a model
   consisting of a straight line in log space and an intrinsic
   dispersion to the luminosities. Our fitting procedure takes into account the
   uncertainties and upper limits on the radio and X-ray data points
   and is intrinsically symmetrical with respect to the two types of
   luminosity. 90 per cent errors on the fitted slopes were estimated
   using a Monte Carlo method. 

Logarithmic slopes for all subsamples are consistent within the
errors, mainly because of the limited statistics. However, we observe a
trend when taking the best fits at their face value (see Table~~\ref{tab:2lumcorr}, where we also
list the intrinsic dispersion as 1-sigma values of Gaussians in log space). The slope of
the best-fit correlation deviates significantly from 1 only when
samples containing more than one class of object are fitted, and it
becomes very close to 1 when samples consisting only of radio
galaxies or core-dominated quasars are fitted (with a possible
flattening for the latter, consistent with what was observed in larger
samples of core-dominated quasars; e.g., \citealt{kemb86,dmw94}).

The dot-dashed line in Figure \ref{fig:2fluxcorr} shows the
linear regressions for the whole sample, while the dashed line
shows the same regression to the sample composed of core-dominated
quasars and radio galaxies. We also show the linear regressions found
when fitting only core-dominated quasars (continuous line). We observe
that lobe-dominated quasars (asterisks) lie above the  correlation
 valid for core-dominated
quasars (see. e.g., \citealt{dmw94,mjh99}). We also notice that 3 radio
galaxies lie in the $L_{\rm X}-L_{R}$ plane above {\em all} correlations,  suggesting a behaviour similar to
lobe-dominated quasars. However, this is not surprising since the radio
galaxies in question are 3C\,6.1,  3C\,220.1 and 3C\,228, for which a simple
power-law spectral model is a good fit to the data. 
It is interesting that the radio galaxies have excess X-ray emission
over an extrapolation from the core-dominated quasars, and may suggest
a difference in the mechanisms  responsible for the jet-related X-ray emission in the two class of objects.

\begin{table*}
\centering
\caption{Correlation results of linear regression for logarithmic
luminosity-luminosity relations. Small $Z$-values mean the correlation
is less significant.}\label{tab:2lumcorr}
\begin{tabular}{lrcrcrc}
\hline
Subsample & Number & Kendall's $\tau$ & Probability & Intercept &
Slope &Dispersion \\
        &       & Z-value &  & &  &\\
\hline
All sources & 19  & 4.42  & 0.00001 &6.3$^{+1.4}_{-4.1}$ & 0.63$^{+0.15}_{-0.06}$&0.38\\
           &     &  &  & &  & \\
Core-dominated & 4  & 1.08 & 0.2786 & 0.4$^{+9.0}_{-9.0}$ & 0.84$^{+0.30}_{-0.30}$&0.04\\
QSO            &     &  &  & &  &\\
           &     &  &  & &  & \\
Radio galaxies  &10 & 2.06 & 0.0397  & -4.6$^{+7.0}_{-10.0}$ & 1.07$^{+0.4}_{-0.4}$  &0.57 \\
           &     &  &  & &  & \\
Core-dominated    &     &  &  & & &  \\
QSO and         & 14 &3.62 & 0.0003 &5.0$^{+1.9}_{-4.5}$ & 0.68$^{+0.12}_{-0.08}$&0.43\\
radio galaxies           &     &  &  & &  &\\

\hline
\end{tabular}
\end{table*}

\begin{figure}
\includegraphics[scale=0.5,angle=0,keepaspectratio]{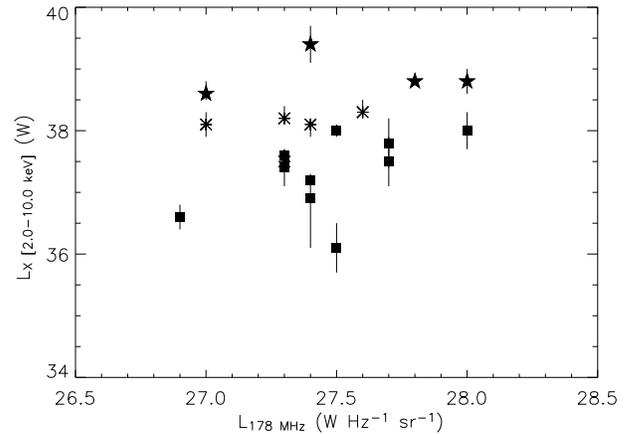}
\caption{X-ray luminosity in the energy band 2.0-10.0 keV (using the
  unabsorbed luminosity of absorbed components in the radio galaxies) versus the 178 MHz
luminosity density. Symbols are as follow: filled squares are radio galaxies; asterisks
are lobe-dominated quasars and stars are core-dominated quasars.
When 1$\sigma$ error
bars are not visible they are smaller than symbols. Axis are the
logarithm of the defined quantity.}
\label{fig:LRLXcorr}
\end{figure}

Absorbed X-ray emission in radio-quiet AGN is associated with the
power of the central engine. We thus also tested possible relations
between the high-energy X-ray emission from the core and the
large-scale radio emission, which should be an indicator of the
overall power emitted over the source life-time.
This is shown in Figure \ref{fig:LRLXcorr}, where the 2-10 keV
luminosity (assuming isotropic emission)  is plotted versus the  178
MHz luminosity density (K-corrected adopting the value of $\alpha_R$ specific to each source
as listed in Table \ref{tab:1kevflux}). For radio galaxies we measured the high energy
luminosity of the absorbed component.
We observe a large dispersion in these two values, showing that if
any correlation exists, it is weak. 

We notice  that radio galaxies lie  below quasars in this plot. This is an additional 
indication that the X-ray emission is relativistically boosted, and
the values for the quasars are higher than those for radio galaxies
due to beaming in the line of sight.  On the other hand, the lack of correlation
between these two parameters may  also be an indication that absorbed
emission from radio galaxies is at least in part jet-related.

\subsection{Spectral index distribution}
\begin{figure}
\begin{centering}
    \includegraphics[scale=0.55,angle=0,keepaspectratio]{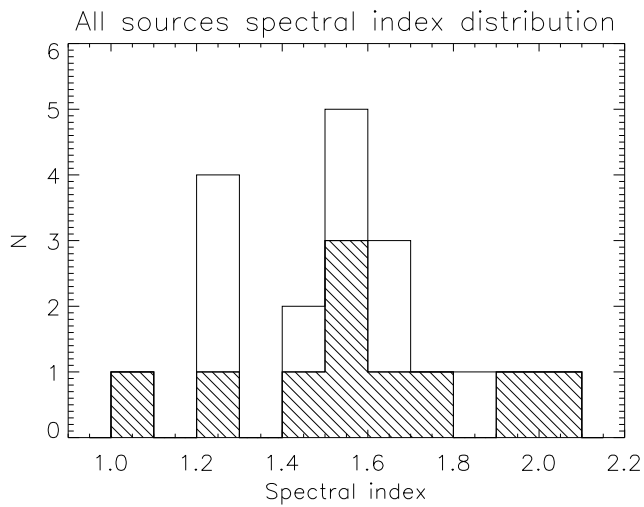}
\caption{Spectral index distribution for the whole sample. The dashed
histogram illustrates the distribution of the absorbed components of
radio galaxies, and empty boxes represent quasars.}
\label{fig:spindexhard}
\end{centering}
\end{figure}

Figure \ref{fig:spindexhard} shows the distribution of the spectral
index for the whole sample. For radio galaxies this is the spectral
index of the absorbed component.
The histogram shows that the population of radio galaxies and quasars
have a similar spectral index distribution. A Kolmogorov-Smirnov
(K-S) test applied to the two populations does not allow us to
  reject the null hypothesis 
 that the two groups of data  are drawn from the same
distribution (the probability of obtaining a K-S statistic equal
  to or more extreme than the one observed is $p = 0.828$).  We calculated mean, median and a bi-square weighted
mean  for the whole sample and each subsample  (see Table
\ref{tab:spindexstat}). In addition to the classical mean value we
calculated a more robust mean using the method described in
\cite{worrallstat}, which takes into account the errors associated with
each data point and fits an intrinsic dispersion in spectral
index. Bearing in mind that the samples are small, the different estimates of the mean values are in agreement. 

The maximum-likelihood analysis gives a  mean spectral index for the whole sample of
1.55$\pm0.17$. Radio galaxies and lobe-dominated quasars have  mean value of $\Gamma$ = 1.57 and $\Gamma=1.59$, respectively,
and the core-dominated quasar subsample has
a mean value of $\Gamma = 1.45$, significantly different from the
other two populations (see Table~\ref{tab:spindexstat}). 

\begin{table*}
\centering
\caption{Average value of spectral index. For radio galaxies both
absorbed and unabsorbed components are tabulated. Col.~1: subsample
under analysis; Col.~2.: number of objects in the subsample; Col.~3: Mean
value and standard deviation; Col.~4: median value and associated
error; Col.~5: bi-square weighted mean and robust error estimate; Col.~6:
Maximum-likelihood method mean and error.}\label{tab:spindexstat}
\begin{tabular}{lccccc}
\hline
Subsample & Number & Mean & Median & BW-mean & ML-mean \\
\hline
All sources & 19  & 1.54$\pm0.19$  & 1.54$\pm0.19$ & 1.54$\pm0.27$ & 1.55$^{+0.05}_{-0.04}$\\
           &     &  &  & &  \\
Absorbed PL  &10 & 1.56$\pm0.21$ & 1.59$\pm0.21$  & 1.56$\pm0.32$ & 1.57$^{+0.03}_{-0.02}$  \\
Radio Galaxies           &     &  &  & &   \\
           &     &  &  & &   \\
Core-dominated & 4  & $1.40\pm0.14$ & $1.54\pm0.14$ & 1.40$\pm0.17$ & 1.45$^{+0.06}_{-0.06}$ \\
QSO            &     &  &  & &   \\
           &     &  &  & &   \\
Lobe-dominated & 5  &1.60$\pm0.19$ & 1.65$\pm0.18$ & 1.60$\pm0.26$& 1.59$^{+0.09}_{-0.09}$\\ 
QSO            &     &  &  & &   \\
           &     &  &  & &   \\
All QSO & 9 & 1.51$\pm0.18$ & 1.54$\pm0.17$ & 1.51$\pm0.23$& 1.52$^{+0.06}_{-0.07}$\\
           &     &  &  & &   \\
Unabsorbed PL&     &  &  & &   \\
Radio Galaxies & 4$^a$  & 2.15$\pm0.28$ & 2.10$\pm0.25$ & 2.13$\pm0.40$& 2.05$^{+0.24}_{-0.25}$\\ 
\hline
\end{tabular}

$^a$ Statistical analysis is performed using only those sources with detections.
\end{table*}

We find a different behaviour when looking at the unabsorbed, soft
power law characterising some of the radio galaxies in the  sample.
For the four radio galaxies for which a spectral index was determined,
we find a mean spectral index $\Gamma_{ML} = 2.05\pm0.25$.  Although errors are large,
this analysis suggests that when an unabsorbed soft component is
present, its spectrum is relatively steep. A K-S test applied to the
spectral index of the 
unabsorbed components of radio galaxies and the absorbed components of
the whole sample allows us to reject the null hypothesis that the
  two samples are drawn from the same population at better than 99 per
  cent confidence.

\section{Discussion}
We have assembled X-ray results for a sample of  19 radio galaxies
(RGs) and
radio-loud quasars (RLQs) for which core spectral properties
have been derived with good precision. Core spectral
properties have been discussed before for RLQs at high
redshift  (e. g.,
\citealt{brinkmann97,brinkmann00, rt00, brunetti02, mjh02, hasen02, donahue03, bel04a,galbiati05}), in
particular for  core-dominated sources with resolved jet-emission
(e.g. \citealt{tavecchio02, gambill03, marshall05}). 

\noindent This is the first time that the
    unabsorbed and absorbed X-ray emission from RGs at $0.5<z<1.0$ has
    been measured with precision in a systematic way and with a
      well-defined sample. Previous large-scale analyses of the X-ray
    spectrum of RGs have been limited to the soft (unabsorbed) X-ray
    emission from these sources and/or have had insufficient spatial
    or spectral resolution to allow separation of core and
    extended emission.

Our sample allows us to discuss our spectral results in the context of
unification schemes.

\subsection{The origin of the core X-ray emission}\label{sec:origin}

The \cha\ spatial resolution allows us to separate emission from the
core and to establish, through spectral fitting, that the emission
from the central 5 pixels of the  observation for 17 of the 18
sources in the sample can be described by one or more  power laws.
3C\,292, observed only with
\xmm, also shows power-law emission in the core
spectrum. The nuclear X-ray
continuum is well described by a single 
power law with Galactic absorption for all
quasars, while 70 per cent of RGs in our sample show intrinsic absorption ranging from
0.3 to 50 $\times10^{22}$ cm$^{-2}$. For about  50 per cent of all sources,
mostly RGs, we 
conclude that a single power-law model does not satisfactorily fit the
data. 

As discussed in the introduction, the  soft X-ray nuclear emission of
core-dominated quasars is consistent with that arising from a relativistic jet, in agreement with results based on data from {\em Einstein} (e.g. \citealt{shastri93}), {\em ROSAT}  (e.g., \citealt{brinkmann97,mjh99}), and ASCA (e.g. \citealt{rt00}).
Only very few radio galaxies have been
detected with {\em ROSAT} at soft energy, and they were found to lie on an
extension of the
flux-flux correlation  for core-dominated
quasars \citep{dmw94, mjh99}. We follow the interpretation of \citet{dmw94} in explaining the lower
X-ray flux density of the radio galaxies as due to the effect of a
relativistic jet which beams the emission from the core-dominated
quasars into the line of sight.

Our study shows that radio-galaxy soft X-ray flux density is indeed
consistent with being unboosted emission from the same component
  as seen in the core-dominated quasars
(Fig.~\ref{fig:2fluxcorr}).
However, the spectrum  is significantly steeper, with an
average value of $\Gamma\sim2.1$ ($\alpha\sim 1.1$), and
is consistent with the  extrapolation of a broken-power law
synchrotron spectrum of $\Delta\alpha\sim0.5-0.9$ from radio
frequencies, as observed in the resolved jets of lower-redshift FRI radio
galaxies (e.g. \citealt{bom872001}, \citealt*{hbw01, dmw01b}). The
steep spectrum disfavours  inverse Compton (IC)
scattering  as the mechanism
for {\it this} emission.
On the other hand, the flat X-ray spectrum ($\alpha\sim 0.45$) observed in
core-dominated quasars is suggestive of IC emission
becoming dominant as the Doppler factor becomes larger (e.g., \citealt{kemb86}).  In
core-dominated quasars the inner jet 
is unobscured, and synchrotron self-Compton emission and upscattering
of external photons from the nucleus are expected to dominate, while
scattering of cosmic microwave background (CMB) photons by a beamed
jet might become important at larger jet radii.
Since increased beaming depresses the X-ray to radio (IC to synchrotron)
ratio,  the slope of the correlation in Figure~\ref{fig:2fluxcorr}
(1.0 for radio galaxies, flattening to 0.8 for
core-dominated quasars) can be qualitatively understood, and confirm
previous work within the limited statistics of our sample of
core-dominated quasars.

The biggest surprise is that, with the exception of 3C\,292 and 3C\,330,
the absorbed  components of radio galaxies  
have flatter spectra ($\Gamma$ = 1.57$^{+0.03}_{-0.02}$)
than those of radio-quiet
quasars (RQQs) (typically $\Gamma\sim$ 1.9-2.0, e.g. \citealt{yuan98,rt00}) and consistent with the
best-fitting power law of RLQs ($\Gamma=1.52^{+0.06}_{-0.07}$), although
somewhat steeper than the core-dominated sample alone ($\Gamma=1.45\pm0.06$).

There are two possible explanations for this observation. One is
    that the X-ray properties of the accretion regions in radio-loud
    and radio-quiet objects are different. However, if we wish to
    retain a simple unification scheme in which radio-loud sources are
    essentially radio-quiet ones with the addition of jets, then our
    results imply that the absorbed emission that we measure in 70 per
    cent of the RGs in our sample is primarily associated
    with the inner jet rather than with the accretion region. There is
    no reason in principle why jet emission cannot come from within
    the torus, and the flat X-ray spectrum of this emission is
    certainly consistent with a jet model.

It is thus interesting to ask whether the central engine is the
    same in radio-loud and radio-quiet objects.
For lobe or core-dominated quasars the beamed
and unbeamed components cannot be separated, but radio galaxies
allow us to try to answer this question. We can ask at what level
absorbed emission with a steep spectrum as measured for RLQs may be present in
radio galaxies, assuming that we are not sensitive enough to measure it.  If a steep
component of high absorption is mixed with  a flatter one with lower
absorption then the fitted slope can be relatively flat and unabsorbed, even if the
absorption-corrected luminosity of the steeper component is significant.

\begin{figure}

\begin{centering}
    \includegraphics[scale=0.55,angle=0,keepaspectratio]{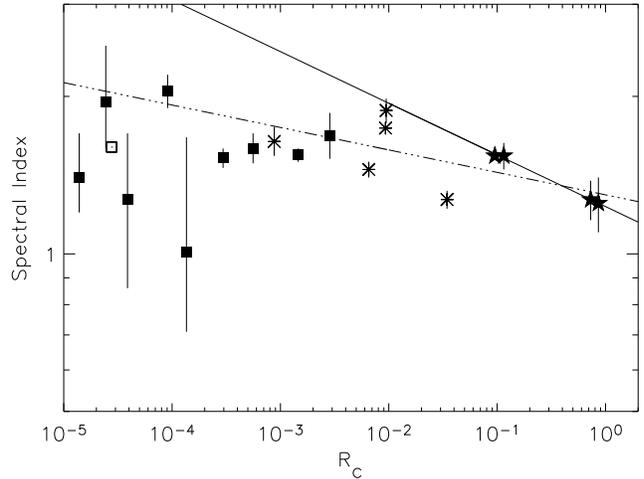}
\caption{X-ray spectral index (absorbed component for radio galaxies) versus the radio core prominence
parameter $R_c$. Symbols
are as in Fig. \ref{fig:2fluxcorr}, the white box is 3C\,427.1 for which the
spectral index was fixed. The dot-dashed line is the best-fit  Buckley-James (B-J) correlation ( as implemented in {\sc asurv} Rev 1.1 and  presented in \citet*{ifn86asurv})
to the subsample of quasars; the continuous line is the best-fit B-J regression
to the subsample of core-dominated quasars.}
\label{fig:spindexvsrc}
\end{centering}
\end{figure}

The result shown in Figure~\ref{fig:LRLXcorr}, in which radio-galaxy
X-ray luminosities are lower than those of quasars, may support this.  Although
we expect core-dominated quasars to have higher luminosities, since no
correction for beaming is applied, this may also suggest that
part of the difference between the two populations of sources arises
because  radio-galaxy luminosities are underestimated if the
correct model has an additional steep absorbed component of high
intrinsic luminosity.

Further support comes from Figure \ref{fig:spindexvsrc}. Here the
spectral index (absorbed component for radio galaxies) is plotted
versus the core-dominance parameter $R_{\rm c}$. This is defined as
the ratio between the core luminosity density at 5 GHz and the large
scale luminosity density at 178 MHz ($R_{\rm c} = L_{\rm 5 GHz} /
L_{\rm 178 MHz}$) and is accepted as an orientation parameter.
In agreement with what has
been found earlier with larger samples (e.g., \citealt{kemb86,
brinkmann97}), at least for core- and lobe-dominated quasars, we
observe a tendency for the spectral index $\Gamma$ to increase with
lower core dominance. However, no
information was available in these earlier studies on the absorbed
components of RGs. 
The sample of RGs discussed here extends
the sampling of the core-dominance parameter by 2
order of magnitude over the use of quasars alone. The
anti-correlation for core-dominated quasars is significant at the 90
per cent level on a Kendall's $\tau$ test. The anti-correlation is
also valid for the population of all quasars (lobe-dominated plus
core-dominated), but the slope tends to zero, the smaller $R_{\rm c}$
becomes. 

This result could be an indication of the presence of an
   additional, {\it isotropic} spectral component in the radio-galaxy
   population. We may be indirectly observing the effects of
   accretion-related emission similar to that observed in RQQs, i.e.
   heavily absorbed and with spectral slope $\sim 2$, which may be
   (partially) masked by a less absorbed, jet-related component.

   In an attempt to test this hypothesis quantitatively for the RGs in
   our sample, we simulated a spectrum in {\sc xspec} using the
response file for the core region of one of the radio galaxies.
We initially simulated a power law of spectral index $\Gamma^{steep} =
1.9$, which is similar to that found for RQQs (e.g. \citealt{rt00})
absorbed by a column density $N_{\rm H}^{steep} = 10^{24}$ cm$^{-2}$,
i.e. $\sim$ a factor of $\sim$10 higher than the average absorption found
for  radio galaxies of our sample. We then added a flatter power law
(of $\Gamma^{flat}=1.57$, the average value for radio galaxies; see Table
\ref{tab:spindexstat}) absorbed by a column density $N_{H}^{flat}$ =
$10^{23}$. For simplicity we ignored a soft unabsorbed component. We varied the normalization of the steeper power
law until the whole spectrum was fitted with a single power law.
We found that with a large number of counts (excellent photon statistics) the
steep power law is washed out if its contribution to the normalization
at 1 keV is $<10$ per cent of the softer power law. 

The statistics in real observations are limited by effective area and 
exposure time. We thus repeated the same exercise
using  photon statistics similar to those of our data.
This yielded the conclusion that a steep, absorbed  power law is not
detected spectrally in the presence of a flatter and less absorbed power law
emission  if the 1-keV  normalizations of the two components are
similar.

If the normalization at 1 keV  of the steeper highly absorbed power law is a factor of two greater
that the original less absorbed  component, then we still find that the whole spectrum
can be fitted with a single power law 
of a spectral index  which is found to be  flatter than $\Gamma^{flat}$. This may explain,
for example, the very flat absorbed-power-law slope of
3C\,265.  The result can be attributed to the ACIS-S response: since the
sensitivity at 1 keV is higher than at higher energies, the effect of
the flat spectrum resulting from the large absorption dominates that
of the steep spectrum at high energy.

A steep-spectrum core-related component of intrinsic
luminosity at least a factor of two greater than  the 2--10 keV
luminosity that 
we actually measure may be present for all our absorbed radio
galaxies, and with the current data we are unable to separate this
component. Our simulations thus support the hypothesis  that the
observed absorbed emission from RGs is jet-related or
jet-dominated. However
we cannot exclude the possibility that the central AGN in RLQs and RQQs
are different.

\subsection{The effect of orientation and implications for unification
schemes}\label{sec:orientation}

\begin{figure}
\begin{centering}
    \includegraphics[scale=0.50,angle=0,keepaspectratio]{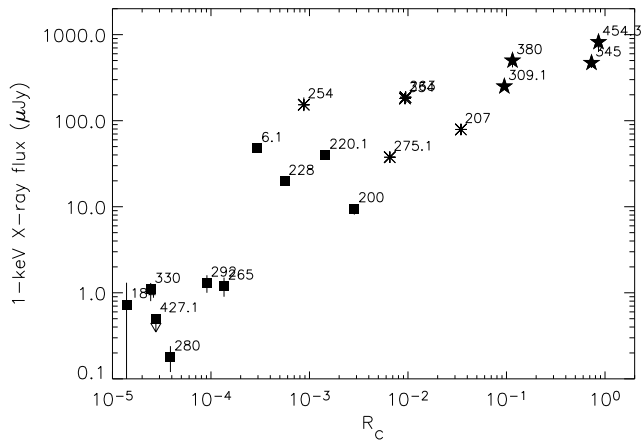}
\caption{X-ray flux density versus the radio core prominence parameter
$R_c$. The 3CRR designation for each galaxy is also marked. Symbols
are as in Fig. \ref{fig:2fluxcorr}.}
\label{fig:1kevfluxvsrc}
\end{centering}
\end{figure}
The correlation between the X-ray flux density at 1 keV and the core prominence
parameter  $R_{\rm c}$ (Fig.~\ref{fig:1kevfluxvsrc})
is an alternative representation of information shown in
Figure~\ref{fig:2fluxcorr}, since the sources have similar isotropic
luminosities and the 5-GHz flux density is beamed.
However, we
notice that the RGs with no or weak
absorption are  those with a larger $R_{\rm c}$ and that they occupy a
position in that plot which is closer to that of lobe-dominated quasars.
This is consistent with the expectations from unified models, in which the X-ray emission from
quasars which are viewed at a large angle to the line of sight is the
sum of jet-related and core-related emission.

The finding that only RGs show intrinsic, rest-frame
absorption is again in  agreement with unification models if these sources are
viewed in the plane of the sky through an absorber. If emission from
RGs is not affected by beaming, we expect a slope close to
1 for the correlation between X-ray and radio core emissions,
which is what we find. Moreover, we also find that the slope of the
correlation valid for radio and X-ray flux density 
becomes smaller the more the beaming is
expected to play a role, as in core-dominated quasars (but note that
this effect is less observed in the luminosity-luminosity relations). This is an
indirect indication of the different effects of beaming in the X-ray and
the radio, which depends on the mechanism responsible for the
emission. We argued in Sec.~\ref{sec:origin} that it is likely that
synchrotron radiation accounts for the jet X-ray emission in radio
galaxies and that IC becomes important, or even dominant, in
  core-dominated quasars.

 We do not find any correlation
between the absorption and the redshift, as found for larger samples of
both RQQs and RLQs (e.g. \citealt{rt00}; but the correlation is
significant only for sources at redshift above 1), but we notice that
the 3 RGs in our sample which do not show intrinsic
absorption are also those at lower redshift.  No
significant correlation was found between the intrinsic $N_{\rm H}$ and the
spectral index, strengthening the constraints we can put on these two parameters,
since they  are intrinsically correlated.

A new result is that
the unabsorbed non-thermal emission from RGs  is consistent with
synchrotron emission from a high-energy extension of the electrons which emit
at radio frequencies. Because of the relatively short energy-loss
timescales of the responsible electrons we may expect soft X-ray
variations on short timescales for at least some of these sources.
This is not observed, but this may be an effect of the 
limited statistics available for such high redshift objects.

\section{Summary and Conclusions}
We have presented X-ray spectra of the unresolved components of 9
quasars and 10 radio galaxies which were similar in isotropic radio
power and which lie in the redshift range $0.5<z<1.0$.  This is the
first time that two spectral components have been reported in the core
spectra of a reasonably large sample of radio galaxies.
We have compared our results with the expectations from a unification model
that describes each source as having a) a central largely  isotropic X-ray
component similar to the X-ray core of a radio quiet quasar; b) a
torus that absorbs X-ray core emission from sources whose jets are in
the plane of the sky (namely radio galaxies), and c) X-ray jet
emission on scales comparable with and larger than the torus

Several results support this simple unification model and allow
conclusions to be drawn about the origin of the X-ray emission in both
radio galaxies and quasars:

\begin{enumerate}
\item Radio galaxies display higher intrinsic absorption than quasars, as
expected from b);
\item An unabsorbed X-ray component is also seen in radio galaxies; we
  interpret this component as arising from the jet. Its steep slope (steeper than the
radio spectrum) is consistent with a synchrotron
origin;
\item Radio galaxies show a radio-core/X-ray-core correlation of slope $\approx1$
when the unabsorbed X-ray component is used. This also argues in
favour of a jet-related emission.
\item The radio-core/X-ray-core correlation flattens when quasars are
considered, i.e. they are relatively X-ray under-luminous for a given radio
flux density. This is qualitatively consistent with IC  X-ray emission
becoming more dominant than synchrotron in the jet as the angle to the
line of sight is decreased and beaming becomes more important. The flat X-ray slope of the
 core-dominated quasars (similar to the radio slope) supports this conclusion.
 Beaming causes the jet to outshine other core components in the X-ray.

\item The lobe-dominated quasars have somewhat more X-ray emission than expected
 on the basis of the jet component in radio galaxies, implying an unabsorbed contribution
 from the inner nucleus that is not outshone by beamed emission in these 
sources. The two contributions to the X-ray emission cannot be
separated with current statistics.

\item  Although the radio galaxies show absorbed X-ray emission, its spectrum is 
flatter than is typically seen in RQQs. This implies that either the
 nuclear emission is affected by the presence of a jet or that the absorbed 
emission is largely from the jet. We have shown through simulations that it is
 possible that a steep-spectrum absorbed core can be present, but masked 
by jet emission.

\end{enumerate}

\section*{Acknowledgements}
E.B. thanks PPARC for support and MJH thanks the Royal Society
  for a research fellowship. We are grateful to D. A. Evans for long and
helpful discussions about pileup, and thank G.W. Pratt for a careful reading
of the manuscript. E. B. thanks M. Cappi for interesting discussion. We
acknowledge an anonymous referee for helpful comments and suggestions
which improved the manuscript. This research has made use of the the SIMBAD database, operated at CDS, Strasbourg, France, and the NASA/IPAC Extragalactic Database (NED) which is operated by the Jet Propulsion Laboratory, California Institute of Technology, under contract with the National Aeronautics and Space Administration.

\appendix

\section{Notes on particular sources and comparisons with the
literature}

\noindent {\bf 3C\,6.1}: the source was observed in two separate
exposures of 36 ks (Obs ID 3009) and 20 ks (Obs ID 4363),
respectively. We extracted the spectrum in the same region (a circle
of radius 2.5 pixels) for the two exposures. Observation 3009 contains
1568 net counts in the 0.2-10.0 keV energy band. Observation 4363
contains 758 counts in the same energy band. The two spectra were
fitted simultaneously. We fitted the spectrum with a model composed of
a power law and a pileup model. However the addition of the pileup
model does not improve the fit, and in each exposure the estimated PUF
is less than 8 per cent. X-ray emission is also observed from hotspots
\citep{mjh04} and from the lobes \citep{croston05}. With  a
mean flux of 0.044$\pm0.008$ for Obs ID 3009 and of 0.039$\pm0.014$
for Obs ID 4363, no significant variability can be claimed.
 
\noindent {\bf 3C\,184}: the \cha\ data do not contain enough counts
(15$\pm4$ net counts) to perform a spectral analysis. For this source
the spectral properties of the core are derived from \xmm\ data. The
source is very small in the radio, with a maximum extent of only 5 arcsec. As discussed in \citet{bel04a} the \xmm\ spectrum  is fitted by a three-component model representing the emission from a cluster-like atmosphere, with best-fit  parameters k$T$ = 3.6$^{+10.6}_{-1.8}$ keV, Z/Z$_{\odot}$ = 0.3, and bolometric $L_{\rm X}= 1.1\times10^{44}$ erg s$^{-1}$; a jet/lobe related soft non-thermal component; and  a core-related absorbed power law component as listed in Table \ref{tab:corehard}.

\noindent {\bf 3C\,200}: the source shows extended X-ray  emission out to 10
arcsec, but point-like emission dominates within the 2.5 pixel radius of the
circle used for the core spectral analysis.  The X-ray properties of
the core of this source have not been described previously in the
literature.

\noindent {\bf 3C\,207}: the core spectrum of 3C\,207 is best-fitted with
a single power-law of rather flat slope $\Gamma=1.27\pm0.05$. This
is in agreement with what was found previously by \citet{brunetti02}
($\Gamma=1.22\pm0.06$), and \citet{gambill03}
($\Gamma=1.36\pm0.06$). We do not find evidence for absorption above the
Galactic value as found by \citet{brunetti02}, and, although we do not
exclude emission of ionized gas in the nucleus, the addition of a
Gaussian iron line does not improve the fit statistically. Hotspot,
jet, lobe and extended emission were also previously discussed
(e.g., \citealt{brunetti02}, \citealt{mjh04}, \citealt{gambill03}).
The 2.0--10.0-keV flux  we found is somewhat lower than found
by \citet{gambill03}, but the two values agree within 20 per cent.

\noindent{\bf 3C\,220.1}: the source shows extended emission which is
likely to be cluster-like in origin \citep{dmw01}. The spectral index
we found is somewhat flatter than that found by
\citet{dmw01}, who  found that intrinsic absorption was
required. When  our results are compared with the Worrall et al. fit
without intrinsic 
absorption  then the two slopes are in
good agreement. Moreover, after taking into account the different  cosmological parameters, fluxes and luminosities are in
agreement.

\noindent {\bf 3C\,228}: the source was observed in two separate
exposures of 10 ks (Obs ID 2453) and 13 ks (Obs ID 2095), respectively,
after flare screening. The shorter observation detected 281 net
counts, while the longer has 365, consistent with no flux variation
between the two.

\noindent {\bf 3C\,254}: the source is piled up. Within uncertainties
our
power-low slope is in  agreement with  that  found by
\citet{donahue03} using the same data. The  flux and luminosity are also in
agreement after considering the different cosmological parameters. The 1-keV flux
density is  1/3 greater than that found by \citet{mjh99}, but this
is likely to be within the large uncertainties arising from the {\it ROSAT}
analysis.

\noindent {\bf 3C\,263}: the source is heavily piled up. We examined
the spectrum extracted in several regions and we finally adopted the
annular region of inner radius 3 pixels and outer radius 5 pixels as
the least affected by pileup. The background spectrum was extracted in
an annulus of inner and outer radii 9 and 12 pixels, respectively. We
also excluded the emission from the jet using a rectangular region.
The power-law slope agrees with that found by \citet{mjh02} using the same data
to within 1$\sigma$ errors. The 1-keV flux density differs
significantly, with our value found to be 50 per cent lower, even
after according a systematic error of 10 per cent to account for the
PSF correction factor. This is likely to be due to a combination of a
different region used (Hardcastle et al. used a circle of radius 5
pixels) and improvement in the calibration since the Hardcastle et al.
analysis, or their use of a pileup model rather than fitting the
spectrum in an annulus.

\noindent {\bf 3C\,265}: the fit of this source spectrum requires 5
components. The model is composed of  Galactic absorption, a power
law, an absorbed power law and a Gaussian at the energy of the neutral
iron line. \citet{bondi04} discussed the properties of this source in
detail.
We find that the slope of the absorbed power law is consistent with the result of \citet{bondi04}.

\noindent{\bf 3C\,275.1}: there is no evidence for X-ray  emission from
a resolved jet, but
the source shows extended emission which has been interpreted 
 as arising from a cluster-like atmosphere \citep{crfab03, mjh00}
 or from lobe inverse-Compton emission \citep{croston05}.

\noindent {\bf 3C\,280}: this source is one of the best studied powerful radio
galaxies. It has an extended emission-line (OII) nebula
with a central peak and a loop around the eastern radio lobes
(e.g., \citealt{mccarthy87, ridgway04}). We selected a
background in the canonical annulus of inner radius 3.5 pixels and
outer radius 6 pixels, but we also excluded the eastern and western
regions corresponding to the X-ray lobes. There are only 50 net source
counts, but these are at energies above 1 keV, suggesting that we are
seeing an absorbed power law. \citet{donahue03}, also using \cha\ data, find a very flat
spectral index, consistent with zero, when fitting between 0.3 and 8.0
keV. The 1-keV flux density we found is in agreement with their
result. 3C\,280 was one of the few high-redshift radio galaxies with
a clear {\it ROSAT} detection \citep{dmw94}. Discrepancies between the {\it
  ROSAT}
and \cha\ results have been discussed by \citet{donahue03}.

Constraints on the soft emission were obtained by fixing the slope of
the unabsorbed power law to 2.1, as for other radio galaxies for which
direct fitting was possible. The absorbed component parameters were
also fixed to obtain constraints on the normalization of the
unabsorbed component. Because parameter values have been fixed, we added a systematic
error of 20 per cent to the flux and luminosity of the unabsorbed
component.

\noindent {\bf 3C\,292}: we used \xmm\ data for this source. Details
of the analysis are described by \citet{bel04a}. The core spectrum
extracted in a circle of radius 10 arcsec was fitted with a
two-power-law model. The source also shows extended X-ray emission
coincident with the radio lobes but little spectral evidence for a cluster-like atmosphere.

\noindent{\bf 3C\,309.1}: our estimate of the X-ray flux density at 1 keV
is in very good agreement with previous results from ROSAT observations
\citep{mjh99}.

\noindent {\bf 3C\,330}: the source X-ray image shows the presence of
some extended emission aligned with the radio lobes (see also
\citealt{croston05, mjh02}). Core
emission was analysed previously using \cha\ data \citep{mjh02} and
 results are in good agreement despite the
slightly different regions used for spectral analysis.

\noindent {\bf 3C\,334}: a reduced $\chi^2$ of 1.35 suggests that an
additional component may be necessary. However, the addition of a
thermal or power-law component does not improve the fit. Some
residuals are observed at low energy (below 0.7 keV) suggesting a soft
excess.

\noindent {\bf 3C\,345}: in agreement with \citet{gambill03} we find that
two power laws are necessary to fit the spectrum from this source: a
steeper ($\Gamma=1.8$) slope fits the soft spectrum, while a flatter
($\Gamma\sim1.3$) is used to represent that hard spectrum. Using
the broken-power-law model we find an energy break of
1.74$\pm0.33$ keV,  in agreement within the errors with
\citet{gambill03}. 

\noindent {\bf 3C\,380}: the strong pileup for this source led us to
mask a central  circle of radius 1.5 arcsec. Our result is in
agreement with that found by \citet{prieto96} using {\it ROSAT} data
($\Gamma=1.58\pm0.13$) and consistent with the photon index  recently published by
\citet{marshall05}: $\Gamma=1.61\pm0.09$.  However there is a
discrepancy between our and their measure of the X-ray flux.
We and \citet{marshall05} extracted the spectrum from the same
area; we used slightly different background, but the background is
negligible for such a high-flux source. However, the big
difference between the two studies is the calibration. Our newer
analysis has taken into account the charge transfer inefficiency and
the ARF is better determined. Moreover, without using a pileup model,
Marshall et al. found a poor fit, with a spectrum that was
inconsistently flatter than that in the current work, and
consequently a lower  1-keV flux density. Marshall et al. attributed
the flat spectrum and bad fit to pileup, but the improved calibration
now available allows a more precise flux density to be determined.

\noindent {\bf 3C\,427.1}: the statistics are not sufficient to perform conclusive
spectral fitting. The spectrum is dominated by hard emission, with
only 3 of the 15 counts found at energy below 2 keV. We fixed the
parameters for an absorbed power law to typical values found for the
other radio galaxies in the sample. We thus adopted a model composed of
an absorbed power law of spectral index 1.6,  and intrinsic absorption
1$\times10^{23}$ cm$^{-2}$. We also obtained an upper limit for the
presence of an unabsorbed soft component.

\noindent {\bf 3C\,454.3}: the source is  heavily piled up. We extracted
the spectrum in an annulus of inner and outer radii 2.0 and 2.5
pixels. Although in this region any effect of pileup is negligible
(and indeed the addition of a pileup model does not improve the fit),
the best-fit spectral index is rather flat. \citet{marshall05} found a
slightly steeper spectral index, but this is attributable to the
different calibration  used (see also 3C\,380). As in the case of 3C\,380, when the spectrum is
fitted above 1 keV the two results are in better agreement. Our
best-fit $\Gamma$ is also in agreement with \citet{tavecchio02} who
used {\it BeppoSAX} data. Discrepancy in the 1-keV flux quoted by us and Marshall et al. are explicable as for the case of 3C\,380.

\end{document}